\documentclass[onecolumn,nobibnotes,nofootinbib]{revtex4}
\usepackage{subfigure}
\usepackage{epsfig}
\usepackage{multirow}
\usepackage{color}
\usepackage{array}
\usepackage{float}
\usepackage{url}
\usepackage{amsmath,amssymb,bm}
\usepackage{appendix}
\newcommand{\be}{\begin{eqnarray}}
\newcommand{\ee}{\end{eqnarray}}

\makeatletter
\newcommand{\thickhline}{%
    \noalign {\ifnum 0=`}\fi \hrule height 1.2pt
    \futurelet \reserved@a \@xhline
}
\newcolumntype{"}{@{\hskip\tabcolsep\vrule width 1.2pt\hskip\tabcolsep}}
\makeatother
\begin{document}
\title{Asymptotic gluon density within the color dipole picture in the light of HERA high-precision data}
\author{D. A. Fagundes} 
\email[E-mail me at: ]{daniel.fagundes@ufsc.br}
\affiliation{Department of Exact Sciences and Education, CEE. Federal University of Santa Catarina (UFSC) - Blumenau Campus, 89065-300, Blumenau, SC, Brazil}

\author{M.V.T. Machado}
\email[E-mail me at: ]{magno.machado@ufrgs.br}
\affiliation{High Energy Physics Phenomenology Group, GFPAE. Institute of Physics, Federal University of Rio Grande do Sul (UFRGS) Postal Code 15051, CEP 91501-970, Porto Alegre, RS, Brazil}

\begin{abstract}
We present an analysis of the most precise set of HERA data within the color dipole formalism, by using an analytical gluon density, based on the double-logarithm approximation of the DGLAP equations in the asymptotic limit of the scaling variable, $\sigma=\log{(1/x)}\log{(\log{(Q^2/Q_ 0^2)})}\rightarrow \infty$. Fits to data, including charm and bottom quarks are performed and demonstrate the efficiency of the model in describing the reduced cross section, $\sigma_{r}$, in the wide range $Q^2:(1.5,500)$ GeV$^2$ for two dipole models including parton saturation effects. We also give predictions to $F_{2}^{c\bar{c}}$ , $F_{2}^{b\bar{b}}$ and $F_{L}$, all describing the data reasonably well in the range $Q^2:(2.5,120)$ GeV$^2$. Total cross sections of exclusive photoproduction of $J/\psi$ and $\rho$ are also calculated and successfully compared to HERA data and recent measurements at LHCb.


\end{abstract}
%
%
\maketitle
%

%
\section{Introduction}
\label{sec:intro}
The QCD color dipole formalism provides an intuitive description of inclusive and exclusive processes in electron-proton ($ep$) and lepton-nucleus ($\ell A$) scattering at high energies. Besides taking into account higher twist contributions beyond the leading-twist approximation it also allows to include corrections related to the parton saturation phenomenon. Namely, the unitarity bounds to scattering amplitude are easily implemented in the dipole approach. 
The key feature is the connection of the dipole-target amplitude, $N (x,r)$, to the integrated gluon density, $xG(x,Q^2)$. The parton saturation models shed light on the behavior of the gluon density at very low-$x$ and this knowledge is crucial for instance to describe the exclusive processes in $ep$ and $eA$ collisions \cite{Morreale:2021pnn}. 
We quote the works in Refs. \cite{Rezaeian:2013tka,Rezaeian:2012ji,Luszczak:2016bxd,Golec-Biernat:2017lfv,Mantysaari:2018nng,Mantysaari:2018zdd,Amaral:2020xqv} where the high quality HERA data are analysed within the dipole model formalism. Notice that the AGBS model \cite{Amaral:2020xqv} is the only one which considers the dipole approach in momentum space. In contrast, analysis using conventional QCD DGLAP formalism presented some tension at small $x$ and $Q^2$ which is mitigated by introducing the $\ln (1/x)$-resummation effects to QCD splitting functions and coefficient functions \cite{Bonvini:2017ogt,Ball:2017otu}. It has been found that those corrections provide an important improvement in the description of the high precision data \cite{xFitterDevelopersTeam:2018hym,Bonvini:2019wxf,Ball:2017otu}.

On the other hand, in inclusive or diffractive DIS at intermediate or large $Q^2$ a precise understanding of gluon density is decisive  because its QCD evolution drives the behavior of the {proton structure functions $F_{2,L}(x,Q^2)$} and the observables characterized by a sufficiently high hard scale.  Therefore, an accurate determination of $xg(x,Q^2)$ is desirable, which can be obtained from an analysis of high precision HERA data on inclusive DIS, available in a wide $Q^2$ region. For low-$x$ and high $Q^2$ it is well known that perturbative QCD accounts for the resummation of the dominant double logarithmic (DLA) contributions $[\alpha_S \ln(1/x)\ln (\mu^2)]^m$. The DLA approximation gives the following scale dependence for the gluon density, $xg(x,\mu^2)\propto \exp [\sqrt{C\ln (x_0/x)\ln (t/t_0)}]$, with $t/t_0 \equiv \ln (\mu^2/\Lambda_{\mathrm{QCD}}^2)/\ln (Q_0^2/\Lambda_{\mathrm{QCD}}^2)$ and $C\simeq  16N_c/\beta_0$. {Here, $N_c$ is the number of colors and $\beta_0 = 11 -\frac{2}{3}N_f$ ($N_f$ is the number of active flavours) is the first term of the QCD beta-function series.} In the general case, the numerical DGLAP evolution at leading order (LO) or next-to-leading order (NLO) is known to describe the gluon distribution at large $Q^2$.
Within the color dipole approach, the impact parameter saturation model (IP-SAT) and the BGK model include DGLAP evolution. In both models the gluon density is parameterized at the initial scale $Q_0^2$ and then evolved to higher
scales $\mu^2$ by using LO or NLO DGLAP evolution codes. 

In Ref. \cite{Thorne:2005kj} a matching between the
dipole model gluon distribution, $\hat{\sigma}(x,r^2)\approx \frac{\pi^2\alpha_s}{N_c}r^2xg(x,\mu^2=\frac{A}{r^2})$, and the collinear DGLAP one is obtained by using  a leading order gluon anomalous dimension $\gamma_{gg}$, which for a flat input at $Q_0^2$ gives the following solution to the $\mu^2$-evolution \cite{Ball:1994du}:
\begin{eqnarray}
xg(x,\mu^2) \propto I_0 \left(2\gamma \sqrt{\ln \left(\frac{x_0}{x} \right)\ln \left(\frac{t}{t_0} \right)}\right)  \exp\left[-\delta \ln \left(\frac{t}{t_0} \right)  \right], 
\label{Thorne:start}
\end{eqnarray}
with $\gamma = \sqrt{12/\beta_0}$, $\delta = (11+\frac{2N_f}{27})/\beta_0$ {and $I_0$ is the modified Bessel function of first kind. The parameter $A$ in $\mu^2=A/r^2$ appears in the ansatz for the relationship between four-momentum scales for the gluon density and transverse dipole sizes, $r$.} A reasonable fit to existing data at that time was obtained with an appropriated extrapolation of the shape in Eq. (\ref{Thorne:start}) to the low $Q^2$ region. However, the conclusion was that the procedure to compute the structure functions
from the dipole cross section does not take into account all contributions which appear in the exact $\alpha_S^n$ perturbative calculations. {In any case,} the ansatz for the gluon density, Eq. (\ref{Thorne:start}), is well founded and takes advantage of the effective asymptotic behaviour of parton densities at any $x$ and $Q^2$. For instance, in Ref. \cite{Ball:2016spl} a methodology has been constructed  to obtain the asymptotic singlet and gluon distributions and their regime of validity. The obtained predictions at large and small Bjorken-$x$ are in good agreement with the updated global parton distribution function (PDF) fits. In particular, at small-$x$ the transition from low $Q^2$ to a high virtuality region is well determined. Starting from a sufficiently soft behavior at the input scale one obtains the double asymptotic scaling for the gluon density \cite{Caola:2008xr,Ball:2016spl}:
\begin{eqnarray}
xg(x,\mu^2) = \frac{A_g}{\sqrt{4\pi \gamma \sigma}}\exp \left[ 2\gamma \sigma - \delta \frac{\sigma}{\rho} \right], \quad \sigma \equiv \sqrt{\ln\left( \frac{x_0}{x} \right) \ln \left( \frac{t}{t_0}  \right) }, \,\,\,\rho \equiv \sqrt{\ln\left( \frac{x_0}{x} \right)/\ln \left( \frac{t}{t_0}  \right)}
\label{eq:model1}
\end{eqnarray}
where the parameters $x_0\simeq 10^{-1}$ and $Q_0^2\simeq 1$ GeV$^2$ set the boundaries of the asymptotic domain {and $A_g$ is the distribution overall normalization}. {The quantities $\sigma$ (geometric mean) and $\rho$ (ratio) are the  double scaling variables \cite{Caola:2008xr,Ball:2016spl}.  Thus, the gluon distribution scales asymptotically in both $\sigma$ and $\rho$ in the double limit of large $\sigma$ at fixed $\rho$ and large $\rho$ at fixed $\sigma$. }

{From the theoretical point of view, it has been demonstrated \cite{Bialas:2000xs} that in the leading-logarithm approximation there is an exact equivalence of $k_T$-factorization formalism (BFKL dynamics) with the color dipole approach. Thus, the high $Q^2$ BFKL limit is given by double-leading-logarithm approximation (DLLA) solution. The DLLA solution is also the common low-$x$ limit of DGLAP evolution equation. Therefore, phenomenological models using full DGLAP evolution in dipole cross-section are using an  educated working hypothesis. This is the reason why we propose to investigate in this paper the DLLA solution for the gluon PDF in the context of color dipole approach. Of course, the extra advantage is the analytical form of such a distribution and very small number of free parameters. The approach considered is completely consistent with the one presented long time ago by Nikolaev and Zakharov \cite{NIKOLAEV1994157} and Thorne \cite{Thorne:2005kj}. Furthermore, a characteristic feature of the double asymptotic scaling (DAS) approximation is that the geometric scaling phenomenon appearing in the photon–proton cross section data is easily understood in the context of DLLA perturbative evolution with generic boundary conditions \cite{Caola:2008xr}.}

It is the purpose of this work to use the well defined asymptotic behavior of gluon density discussed above in the context of the parton saturation models. This replaces the usual (numerical) LO DGLAP evolution in the large virtualities domain by the analytical expressions from pQCD double asymptotic scaling approximation (DAS). In particular, we consider the available models for the dipole cross section which incorporate the evolved gluon PDF. This is the case for the BGK \cite{Bartels:2002cj,Rezaeian:2012ji,Luszczak:2016bxd} and IP-sat \cite{Kowalski:2003hm,Mantysaari:2018nng,Mantysaari:2018zdd} saturation models. {In the context of the present study, BGK and IP-sat models are compared in order to understand the role played by different procedures for implementing unitarity corrections to the bare input involving the gluon distribution.} This paper is organized as follows. In Sec. \ref{sec:2} we describe the $\gamma^*p$ cross section in terms of the dipole scattering amplitude including QCD evolution by mean of the DAS approximation. In Sec. \ref{sec:3} fitting methods to high precision HERA data on the reduced cross section, $\sigma_r(x,y,Q^2)$, are presented along with the fit-tuned parameters to $F^{c\bar{c}},F^{b\bar{b}} $ and $F_L$ structure functions. In Section \ref{sec:4} we extend the formalism to calculate the photoproduction cross section of vector mesons, $J/\psi$ and $\rho$ within the color dipole framework. In the last section, main results are discussed and prospects of possible future studies are presented. 

\section{Theoretical framework and phenomenological models}
\label{sec:2}
Before proposing the models for dipole cross section with QCD evolution inspired by the DAS approximation, we present a short review of the application of the dipole formulation of the photon-nucleon scattering. The cross section for the interaction of a virtual photon with a given polarisation (transverse, $T$, or longitudinal, $L$) off a proton target is expressed as:
\be
\label{sigtot}
\sigma_{T,L}^{\gamma^*p}(x,Q^2)= \sum_f
\int d^2\vec{r}\int dz \left|\psi_{T,L}^f(Q,r,z)\right|^2 \hat\sigma (x,\vec{r}), 
\ee
where $\psi_{T,L}^f(Q,r,z)$ is the corresponding photon wave function in mixed representation and $\hat\sigma (x,r)$ is the dipole cross section. The label $f$ refers to the quark flavour. 
{A tacit ad hoc assumption is that the dipole cross section depends on Bjorken-$x$ variable instead of $W$, the photon-proton centre of mass energy. This procedure is widely used and yields acceptable fits to the experimental data. It can be traced back to the original works on QCD dipole picture that the dependence on $W^2$ is associated to the 
life-time of a $q\bar{q}$ fluctuation that is the space-time interpretation of Generalized Vector 
Dominance (VDM) and implies its dependence of $\hat\sigma$ (see discussion on Refs. \cite{Ewerz:2004vf,Ewerz:2006vd,Ewerz:2011ph}). There are a few models incorporating the $W$ dependence in the dipole-cross section. Some examples are the Donnachie-Dosch \cite{Donnachie:2001wt}, Forshaw-Sandapen-Shaw \cite{Forshaw:2006np} and Schildknecht \cite{Schildknecht:2020oug,Kuroda:2017ogq} models. They are able to describe both proton structure functions as exclusive vector meson production \cite{Forshaw:2006np,Forshaw:2003ki,Schildknecht:2016jqa}. Unfortunately, most of them have not been updated by using the high precision data.
}

The squared photon wave functions summed over the quark helicities for a given photon polarisation and quark flavour $f$ are expressed by 
\begin{eqnarray}
 \left|\psi_{T}^f(Q,r,z)\right|^2 & = &  \frac{2N_c}{\pi}\alpha_{\mathrm{em}}e_f^2\left\{\left[z^2+(1-z)^2\right]\epsilon^2 K_1^2(\epsilon r) + m_f^2 K_0^2(\epsilon r)\right\},  \\
  \left|\psi_{L}^f(Q,r,z)\right|^2 & =& \frac{8N_c}{\pi}\alpha_{\mathrm{em}}e_f^2 Q^2 z^2(1-z)^2 K_0^2(\epsilon r).
\end{eqnarray}
where $K_{\nu}$  are the modified Bessel functions of second kind of order $\nu=0,1$ and $\epsilon = \sqrt{z(1-z)Q^2+m_f^2}$.

Given the conservation of the dipole transverse size $\vec{r}$ during the collision the dipole formula can be related to the unintegrated gluon distribution (UGD), ${\cal{F}}(x,\vec{k})$, in  the $k_{\perp}$-factorization approach, 
\be
\label{UGDdipole}
\hat\sigma(x,\vec{r})= \frac{2\pi}{3}\int \frac{d^2\vec{k}}{k^4}\ \alpha_S {\cal{F}}(x,k^2) (1-e^{i\vec{k}\cdot \vec{r}})(1-e^{-i\vec{k}\cdot \vec{r}}).
\ee
In the color transparency domain, $r\rightarrow 0$, the dipole cross section is related to the gluon density \cite{Blaettel:1993rd,Frankfurt:1996ri},
\be
\label{CT}
\hat\sigma(x,r) \simeq
\frac{\pi^2}{3}r^2\alpha_S xg(x,\mu^2),
\ee
where the scale is set $\mu^2 = C/r^2$. In the BGK saturation model \cite{Bartels:2002cj,Rezaeian:2012ji,Luszczak:2016bxd}, the dipole cross section incorporates the evolved gluon density. It is evolved with the LO or NLO DGLAP evolution equation and neglecting  quarks in the evolution. Now, the scale takes the form $\mu^{2}=\frac{C}{r^2}+\mu_0^2$, where the parameters $C$ and $\mu_0^2$ are determined from a fit to DIS data.
\be
\label{BGK}
\hat\sigma (x,r)=\sigma_0\left\{
1-\exp\left(-\frac{\pi^2 r^2 \alpha_S(\mu^2)\,xg(x,\mu^2)}
{3\sigma_0}\right) \right\}.
\ee
The gluon density is parametrized at the starting scale $\mu_0^2\sim 1$ GeV$^2$. {The parameter $\sigma_0$ is related to the target transverse area, $\sigma_0\approx 2\pi R_p^2$}. The quantity $x$ used in expressions above is the modified Bjorken variable, $x=x_{\mathrm{Bj}}\left(1+\frac{4m_q^2}{Q^2}\right)$, with $m_q$ being the effective quark mass. {Such replacement was first proposed in the celebrated GBW papers \cite{Golec-Biernat:1998zce,GolecBiernat:1999qd} and considered for the majority of color dipole models in literature since then. The main point is to describe consistently the transition from high $Q^2$ towards the photoproduction limit $Q^2\rightarrow 0$.} This modification is quite important as heavy quarks contribution are taken into account. The soft ansatz (S) as in the original BGK model \cite{Bartels:2002cj,Luszczak:2016bxd,Golec-Biernat:2017lfv}  and the updated soft + hard  ansatz (S+H) \cite{Luszczak:2016bxd} take the form:
\be
\label{eq:gluon}
xg_{\mathrm{S}}(x,\mu_0^2) & = &  A_g x^{-\lambda_g} (1-x)^{C_g}, \\
xg_{\mathrm{S+H}}(x,\mu_0^2) & = & A_g x^{-\lambda_g} (1-x)^{C_g}
(1 + D_gx + E_gx),
\ee
where a fixed parameter $C_g = 5.6$ is frequently used and $A_{g}, \lambda_g, D_g$ and $E_g$ are free fit parameters. The saturation scale, $Q_s$, which defines the transition to the saturation region  is usually obtained from the condition $\frac{4\pi^2}{3\sigma_0Q_s^2}xg(x,\mu_s^2)=1$ with $\mu_s^2 = \frac{C}{4}Q_s^2+\mu_0^2$.

The impact-parameter saturation model (IP-sat) \cite{Kowalski:2003hm,Mantysaari:2018nng,Mantysaari:2018zdd} is based on the Glauber-Mueller dipole cross section and describes the interaction of a QCD dipole probe with a dense target. The $b$-dependence of the dipole cross section is crucial to describe the momentum transfer  $|t|$ distributions in exclusive processes like vector meson production and deeply virtual Compton scattering (DVCS). The dipole cross section is obtained from S-matrix element at a given $\vec{b}$,
\be
\hat \sigma(x, \vec{r}) &=&\int d^2\vec{b}\, \frac{d\sigma_{q\bar{q}}}{d^2\vec{b}}= \int d^2\vec{b} \,\,2 \left(1 -  \mbox{Re}\, S(b)\right), \\
\frac{d\sigma_{q\bar{q}}}{d^2\vec{b}} &=& 2\,\left[1-
\exp\left(-\frac{\pi^2}{2\,N_c}r^2\alpha_S(\mu^2)xg(x,\mu^2)T(b)\right)
\right].                 
\label{IP-satDIP}
\ee
In general, a gaussian form is taken for the proton thickness function $T(b)$ which is motivated by the $|t|$-distribution in quarkonia exclusive production \cite{Cepila:2019skb,Henkels:2020kju,Henkels:2020qvo,Jenkovszky:2021sis}. The proton profile function properly normalized is written as:
\be
\label{TBPROTON}
T_G(b) = \frac{1}{2\pi B_{G}} \exp \left(-\frac{b^2}{2B_{G}}\right), \quad  \int d^2\vec{b}\,T_G(b) =1,
\ee
where the parameter $B_G\simeq 4$ GeV$^2$ is related to the average squared transverse radius of the nucleon, $\langle b^2  \rangle = 2B_G$,  and to the charge proton radius, $R_p=\sqrt{2B_G}$ \cite{Kowalski:2003hm,Mantysaari:2018nng,Mantysaari:2018zdd}. The saturation scale with $b$-dependence is defined as:
\be
Q_s^2(x,b) = \rho (x,R_s,b), \quad \rho (x,r,b)\equiv \frac{2\pi^2}{N_c}\alpha_S(\mu^2)xg(x,\mu^2)T(b),
\ee
where $\mu^2 = \mu^2 (r)$ and the saturation radius $R_s=R_s(x,b)$ is given by the numerical solution of the transcendental equation $\rho (x,R_s,b) = 2/R_s^2$. In Ref. \cite{Mantysaari:2018nng} a linearized version of the IP-sat model has been investigated. In this case, the large dipole contribution is controlled by requiring a sort of  confinement effect suppressing dipoles larger than the inverse of light quark mass. Moreover, the IP-sat saturation model is the basis for the Sar$t$re Monte Carlo \cite{Toll:2013gda}.

The sucess of the models with DGLAP evolution discussed above motivate the construction
of an analytical expression for the dipole cross section based on the DAS approximation.  A clear advantage is that numerical evolution codes are not needed in order to evolve the gluon distribution to the hard $\mu^2$ scale. We also explore the implications of the analytical model to the description of structure functions.  This will be done
in what follows.

\section{Fit procedures and results}
\label{sec:3}
\subsection{Tests of the original DAS for the gluon distribution}
The first improvement we make in this study, at least in comparison with Thorne's model \cite{Thorne:2005kj}, is the use highest-precision HERA data \cite{Abramowicz:2015mha}, including heavy -- charm and bottom -- quarks, in the fits with the gluon PDF from the DAS solution. In particular, we fit the reduced cross section data \cite{Abramowicz:2015mha}, which reads
\begin{equation}\label{eq:red_xsec}
\sigma_{r}(x,y,Q^{2})=F_{2}(x,Q^{2})-\frac{y^{2}}{1+(1-y)^2}F_{L}(x,Q^{2}).
\end{equation}
where $y = Q^2/(sx)$ is the inelasticity variable, $\sqrt{s}$ denotes the center of mass energy of the $ep$ collision, $F_{T/L}(x,Q^{2})$ correspond to the transverse/longitudinal structure functions and $F_{2}(x,Q^{2})=F_{T}(x,Q^{2})+F_{L}(x,Q^{2})$. 

To start off, we select a narrow bin with $Q^{2}:(1.5,50)$ GeV$^2$ and $x<0.001$ ($N=290$) as a conservative choice to test the model of Eq. (\ref{eq:model1}) at low$-x$ values and intermediate $Q^2$; a kinematical window for which we did expect the model to work, just as in the original tests of the gluon DAS solution \cite{Ball:1994du}. Moreover, other recent phenomenological studies show its applicability in exclusive $J/\Psi$ \cite{Flett:2020duk} and $\Upsilon$ photoproduction \cite{Flett:2021fvo}.  In this very first test we only include light quarks, while varying the mass $m_{lq}=30-140$ MeV. The results of these initial tests are given in Table \ref{tab:Model1_bin1}.

All fits have been performed using the ROOT framework \cite{Brun:1997pa,Antcheva:2011zz}, through the members of the TMINUIT class. In specific, we use the MIGRAD minimizer and HESSE to check the error matrix wether full convergence has been achieved or only an approximate minimum is found. Also, we have set the confidence level (CL) of our fit parameter to 70$\%$ level throughout.  We also provide the integrated probability, $P(\chi^{2}; \text{d.o.f.})$, the well-known $p-$value, also as a goodness-of-fit estimator, limiting to interpret its results in the light of an overall agreement with datasets for the  models tested.

\begin{table}[H]
\centering
\caption{Free-fit parameters of dipole models BGK and IP-sat assuming the gluon density, $xg(x, \mu^2)$, of Eq.(\ref{eq:model1}) obtained for the bin $Q^{2}:(1.5,50)$ GeV$^2$ and $x<0.001$, with only light quarks included. All fit parameters are given within $70\%$ of confidence level, with $\Lambda=244$ MeV, $\mu_{0}^{2}=1.1$ GeV$^2$, and $C=4.0$ fixed throughout.}
\vspace{.3cm}
\resizebox{.9\textwidth}{!}{
\begin{tabular}{c|c|c|c|c|c|c|c}
\thickhline
Model& $m_{lq}$ [GeV]& $\sigma_0$ [mb]  & $B_{G}$ [GeV$^{-2}$]  & $A_{g}$ & $x_{0}$  & $\chi^2/$dof & $p-$value\\
\thickhline
\multirow{2}{*}{BGK}  & 0.03 & $94.0\pm 7.5$  & \multirow{2}{*}{$-$}  &  $ 1.4061 \pm 0.0074$&  $ 1.00 \pm 0.10$ &  $320.3/287=1.1$ &  $0.086$ \\
& 0.14 &$990 \pm 590$  & &  $ 1.4209 \pm 0.0082$ & $ 0.998 \pm 0.054$  & $381.5/288=1.3$  & $1.5\times 10^{-4}$\\
\hline
\multirow{2}{*}{IP-sat}  & 0.03  &  \multirow{2}{*}{$-$}  &  \multirow{2}{*}{4.6 (fixed)}  & $  1.614 \pm  0.028$ &  $ 0.681 \pm 0.046 $ &  $326.5/288=1.1$ &  $0.059$ \\
& 0.14 & &  &  $ 1.845 \pm 0.050$  & $ 0.484 \pm 0.050$ & $409.9/288=1.4$  & $3.0\times 10^{-6}$\\
\thickhline
\end{tabular}
}
\label{tab:Model1_bin1}
\end{table}

From these results one first noticed a preference for smaller light quark masses, as the best fits are obtained with $m_{lq}=30$ MeV. In addition to that, fits comprising only light quarks produce higher values of $\sigma_{0}$ in comparison, for instance with GBW2018 \cite{Golec-Biernat:2017lfv} which finds $\sigma_{0}\sim 23$ mb in the version including only light flavours (c.f. Table II of Ref. \cite{Golec-Biernat:2017lfv}). Moreover, fits with BGK clearly favours $x_{0}\sim 1$, i.e one order of magnitude higher than usual choice $x_{0}=0.1$ \cite{Ball:2017otu}. Finally, when testing the asymptotic model for the gluon with the BGK dipole ansatz, still with only light quarks included, in a wider $x$ range, namely $x<0.01$, yields $\chi^2/$dof $\sim 1.3$ and $p \sim 10^{-6}$. On the other hand, when heavy quarks are included, the fit quality worsens and we obtain $\chi^2$ values one order of magnitude higher (tipically $\sim 14$). Notwithstanding, we also find very high values of $\sigma_{0}$ for BGK, ranging from typically $10^{2}-10^{7}$ mb, depending on whether heavy quarks have been added or not. 

All theses results together evidence that, extending the model to a higher $x_{max}$ threshold, while including heavy quarks masses in the amplitude, require some modification in the gluon density, to better accommodate larger-$x$ data, while setting $m_{lq}=0.03$ GeV (i.e. the lowest value we have tested). That being put, we set $x_{0}=1$ in the following, while proposing a new model for $xg(x,\mu^2)$ (as we shall fit HERA data with $x<0.01$), firstly identifying a better choice to the  $x$-dependent normalisation at the initial scale $\mu^{2}_{0}$. 

{ At this stage, the quality of the fit can be compared to previous studies in literature. Let us start with the BGK models. In Ref. \cite{Luszczak:2016bxd}, the fit without valence quarks produces a high $\sigma_0\sim 100$ mb by using the soft ansatz of Eq. (\ref{eq:gluon}) and a goodness-of-fit similar to ours. The main difference in the procedure is the inclusion of heavy quarks (charm) and fixed $m_q=0.14$ GeV. The large $\sigma_0$ value thus yield a potential trouble as the  extrapolation of the model to the photoproduction cross section would produce significantly higher than the measured value 174 mb at $W = 209$ GeV (DESY-HERA). Such a large value could also be troublesome for the
description of the inclusive DIS diffractive cross section, which is
more sensitive to large dipole sizes. On the other hand, in the work \cite{Golec-Biernat:2017lfv} a smaller $\sigma_0\sim 23 $ mb is found, given that massless light quarks, $m_q=0$, are considered and along with the inclusion of heavy quarks (charm and bottom). There, the goodness-of-fit is improved by using a different choice of the scale $\mu^2=\mu_0^2/[1-\exp(-\mu_0^2r^2/C)]$. In this context, the fit results presented in Table \ref{tab:Model1_bin1} are quite strict compared to those in \cite{Luszczak:2016bxd,Golec-Biernat:2017lfv} as only three parameters are fitted. In fact, as $x_0$ in  Table \ref{tab:Model1_bin1} is consistent with unity a reasonable fit quality could be achieved by using only two free parameter, $A_g$ and $\sigma_0$. Concerning the IP-Sat model, our results can be compared to the ones in the work \cite{Mantysaari:2018nng}, which include heavy quarks. The goodness-of-fit is also similar to ours, for the same $Q^2$ bin, but in our case by using only 2 free parameters and just light quarks. As already mentioned the introduction of heavy quarks degraded the $\chi^2/\mathrm{dof}$. For completeness, in Appendix A the results of the fits including heavy quarks are presented (for $Q^2:(1.5,50)$ GeV$^2$ and $x<10^{-3}$). The BGK model has been investigated using $m_c=1.3$ GeV and $m_b=4.2$ GeV, respectively. We have also checked that the goodness-of-fit is strongly degraded if the typical value  $x_0\simeq 1$ obtained in the light quark fit is considered. Although, good values of $\chi^2/\mathrm{dof}$ are achieved by using fixed $x_0=0.1$, but still the $\sigma_0$ is very large. Finally, the parameters $x_0$ and $\mu_0^2$ have also been taken free and in such case similar fit quality was achieved by using only 3-4 parameters.}

\subsection{The new DAS inspired gluon PDF}
To account for heavy quark effects in both dipole models, the first modification of Eq.(\ref{eq:model1}) we shall make regards the use of a soft ansatz for the gluon at the scale $\mu^{2}_{0}$, namely:
\begin{equation}
    xg(x,\mu^{2}_{0})=A_{g}x^{-\lambda_g}(1-x)^{C_g},
    \label{eq:gluon_initial}
\end{equation}
where $A_{g}$ and $\lambda_g$ free parameters and $C_{g}=6.0$. That suitable choice has been proven efficient in many phenomenological studies where DGLAP dynamics is tested against the saturation hypothesis \cite{Rezaeian:2013tka,Rezaeian:2012ji,Luszczak:2016bxd,Mantysaari:2018nng,Mantysaari:2018zdd}. Moreover, as we take into account heavy quark contributions we perform the usual kinematic shift in the definition of Bjorken-$x$ \cite{GolecBiernat:1999qd}
\begin{equation}\label{eq:modif-x}
 \widetilde{x}_{f}=x\left(1+\frac{4m_f^2}{Q^2}\right).\nonumber
\end{equation}
for charm and bottom, whenever the cut $\widetilde{x}_{f}\leqslant 0.1$ is satisfied.  In such case, the gluon density is computed at $\tilde{x}_{f}$, otherwise the contribution of heavy quarks is switched off. For light quarks the shift has a negligible effect, thus for light flavours ($u,d,s$) we evaluate  $F_{2}(x,Q^{2})$ and other structure functions at the standard Bjorken-$x$.  

Additionally, to allow a smooth transition from large-to-lower $x$ in this new analytical gluon model, we shall also modify the exponential term present in asymptotic solution of Eq. (\ref{eq:model1}). Such change is motivated by the introduction of massive heavy quarks in the dipole wave functions, which distorts the normalisation of gluon density as well as the transition from higher to lower $x$, during the fitting process. This effect is widely recognized in the literature by the enhancing $\chi^{2}/$dof in fits of HERA data including charm and bottom quarks.

To handle all this aspects, we introduce a modified version of the double-log solution given in Eq. (\ref{eq:model1}).  Namely, we have considered the form,
\begin{align}
xg(x,\mu^2) = xg(x,\mu^{2}_{0})\exp{(-\delta \sigma/\rho)}\exp(2\gamma \sigma')
\label{eq:model2.2}
\end{align}
where $xg(x,\mu^{2}_{0})$ stand for the soft gluon \textit{ansatz} of Eq. (\ref{eq:gluon_initial}) and  
\begin{align}
 \sigma'&= \sigma \sqrt{\mathcal{N}(1-x)^{5}},
\label{eq:newSigma}\\
\rho'&= \rho \sqrt{\mathcal{N}(1-x)^{5}},
\label{eq:rho_new}
\end{align}

in which $\mathcal{N}$  represents a new free parameter of the model that controls the normalization of $xg(x,\mu^2)$ with the evolution of $\mu^2(r)=C/r^{2}+\mu_{0}^{2}$. Such factor corrects the overall gluon normalization as well as the transition to larger $x$ in the asymptotic term $e^{2\gamma \sigma}$ within the original formula. Moreover, keeping 5 active flavours yields $\gamma(n_{f}=5)=1.25$ and $\delta(n_{f}=5)=1.48$ and $\mathcal{N}\simeq 0.2-0.3$. As we shall demonstrate, Eq. (\ref{eq:model2.2}) captures all the essential features of the standard (numerical) DGLAP evolution of gluon in the fits of HERA data \cite{Abramowicz:2015mha,H1:2018flt}. 

Altogether the dipole models we have analyzed comprise 3 to 4 free parameters. IP-sat parameters are only those of the gluon, namely $A_{g}$, $\lambda_{g}$ and $\mathcal{N}$, as we kept $B_{G}=4.0$ GeV$^{-2}$ fixed. On the other hand, BGK requires an extra parameter, the cross section $\sigma_{0}$, that rules the size of saturation effects. {The gluon PDF proposed in Eq. (\ref{eq:model2.2}) has the very same functional behavior at $x\rightarrow 0$ like the gluon distribution considered for instance to describe exclusive $J/\psi$ photoproduction at the LHC within the  $k_T$-factorization approach in Ref. \cite{Jones:2016icr,Flett:2020duk},
\begin{eqnarray}
xg(x,\mu^2) = Nx^{-a}\left( \frac{\mu^2}{Q_0^2}\right)^b\exp\left[\sqrt{16(N_c/\beta_0)\ln (1/x)\ln (t/t_0)}\right],
\end{eqnarray}
with $\Lambda_{\mathrm{QCD}} = 200$ MeV and $Q_0 = 1$ GeV. There, the gluon PDF is fitted by using 3 free parameters ($N,\,a,\,b$) against both HERA and LHCb data for exclusive $J/\psi$ photoproduction at 13 TeV \cite{Jones:2016icr}. In the context of NLO collinear approach the model for the gluon PDF above was also used to compute the cross section for $D$-meson production in the forward direction measured by the
LHCb collaboration \cite{DeOliveira:2017cuj}.}

\subsection{Fits with the new gluon density}

The initial tests of our new gluon model of Eqs. (\ref{eq:model2.2},\ref{eq:newSigma}) consist in finding  an optimal $Q^2$ bin to tune free parameters of the model, namely $A_{g}$, $\lambda_{g}$, $\mathcal{N}$ and $\sigma_0$ (whenever the case). Our procedure have been to look for the widest $Q^{2}$ range with available data yielding the best fit-quality estimators, $\chi^2$/dof and $p-$value, while fixing $Q^{2}_{min}=1.5$ GeV$^{2}$ to consistently test the model in a perturbative region. The results of this $\chi^2$ scanning is shown in the Table \ref{tab:MapChi2_q2max} for model BGK, from which we have set the optimal bin $Q^2:(1.5,50)$ GeV$^{2}$ to our study. {Here, $N$ is the number of data points corresponding to the bin $Q^2:(Q_{min}^2,Q_{max}^2)$. } That put, the free parameters of dipoles BGK and IP-sat have been tunned to fit HERA data in the optimal bin for two datasets: (I) comprising only $\sigma_{r}$ data from Ref. \cite{H1:2015ubc} ($N_{1}=414$) and (II) including $\sigma_{r}^{c\overline{c}}$ data from Ref. \cite{H1:2018flt} in the ensemble I ($N_{2}=414+34=448$).

\begin{table}[H]
\centering
\caption{Fit quality and cross section $\sigma_0$ variation of the BGK model corresponding to changing $Q^{2}_{max}$ in the bin: $(Q^{2}_{min},Q^{2}_{max})$. In all cases we have set $Q^{2}_{min}=1.5$ GeV$^2$. }
\vspace{.3cm}
\resizebox{0.6\textwidth}{!}{
\begin{tabular}{c|c|c|c|c}
\thickhline
$Q^{2}_{max}$ [GeV$^{2}$] & $N$ & $\sigma_{0}$ [mb]& $\chi^{2}/$dof & $p-$value \\
\hline
50 & 414 & $27.3 \pm 6.1$ & $417.468/410=1.02$ & $0.389$\\
\hline
150 & 549 & $25.2 \pm  1.3$& $589.425/545=1.08$& $0.0916$ \\
\hline
250 & 586 & $23.05\pm 0.71  $& $651.486/582=1.12$& $0.0238$ \\
\hline
650 & 620 & $21.65 \pm 0.94 $& $697.836/616=1.13$& $0.0121$ \\
\thickhline
\end{tabular}
}
\label{tab:MapChi2_q2max}
\end{table}

In Table \ref{tab:Model2_bin1} we display the best-fit parameters achieved in the fits to HERA data with BGK and IP-sat, using datasets I and II in the optimal bin previously established, and the gluon from eq. (\ref{eq:model2.2},\ref{eq:newSigma}). As these results evidence, despite the very small uncertainties in dataset I, both models provide very good fits to the data, once $\chi^{2}\sim 1.0$ and $p\sim 0.1-0.3$,  which indicates a very good match between data and models. On the other hand, including charm cross section data worsen the fits, as $\chi^{2}$ increases  by some $20-25\%$, while the $p-$value decreases drastically (at least by 4 orders of magnitude). Despite that, the small variation of fit parameters (by a few percent) from fits to dataset I to II show a good stability of the gluon model, while evidence the difficulty of simultaneously fit $\sigma_{r}$ and $\sigma_{r}^{c\overline{c}}$.  It is worthy noticing that previous analysis of HERA high-precision data, performing numerical DGLAP evolution \cite{Mantysaari:2018nng,Mantysaari:2018zdd} have found very similar results in respect to $\chi^{2}$ enhancement when charm cross section data from \cite{H1:2018flt} is included in the fits.

\vspace{-0.5cm}
\begin{table}[H]
\centering
\caption{Free-fit parameters of dipole models BGK and IP-sat, with gluon density of Eq.(\ref{eq:model2.2}) obtained for the bin $Q^{2}:(1.5,50)$ GeV$^2$ in the datasets
I  and II for $x\leq 0.01$. Charm and bottom quarks are included for $\tilde{x}_{c},\tilde{x}_{b}\leq 0.1$. All fit parameters are given within $70\%$ of confidence level, with $\mu_{0}^{2}=1.1$ GeV$^2$, $C_{g}=6.0$, $C=4.0$, $m_{lq}=0.03$ GeV, $m_c=1.3$ GeV and $m_b=4.2$ GeV fixed throughout.}
\vspace{.3cm}
\resizebox{1.0\textwidth}{!}{
\begin{tabular}{c|c|c|c|c|c|c|c|c}
\thickhline
Model & Dataset &$\sigma_0$ [mb] & $B_{G}$ [GeV$^{-2}$]  & $A_{g}$ & $\lambda_{g}$ &  $\mathcal{N}$ & $\chi^2/$dof & $p-$value \\
\thickhline
\multirow{2}{*}{BGK} &  I &  $27.3 \pm 6.1$ & \multirow{2}{*}{$-$}  & $ 1.210 \pm 0.093$  & $ 0.134 \pm 0.020$  & $0.207 \pm 0.045$ & $417.468/410=1.02$ &  $0.389$\\
& II & $29.2 \pm 2.4$   &   &  $ 1.186 \pm 0.027$  &  $0.1307 \pm 0.0049$  & $ 0.2149 \pm 0.0091$   & $569.863/444=1.28$  & $4.78\times 10^{-5}$ \\
\hline
\multirow{2}{*}{IP-sat} & I & \multirow{2}{*}{$-$}  & \multirow{2}{*}{4.0(fixed)}  &   $ 1.048\pm 0.024$ &  $  0.1017 \pm 0.0053$ &   $  0.2891 \pm 0.0053$ & $447.453/411=1.09$ & $0.104$\\
& II&  &   &  $1.0456  \pm  0.0079$ &  $ 0.1030 \pm 0.0015 $ &  $0.2867 \pm 0.0029 $   & $586.188/445=1.32$ & $7.45\times 10^{-6}$ \\
\thickhline
\end{tabular}
}
\label{tab:Model2_bin1}
\end{table}

\begin{figure}[H]
    \centering
    \includegraphics[scale=0.85]{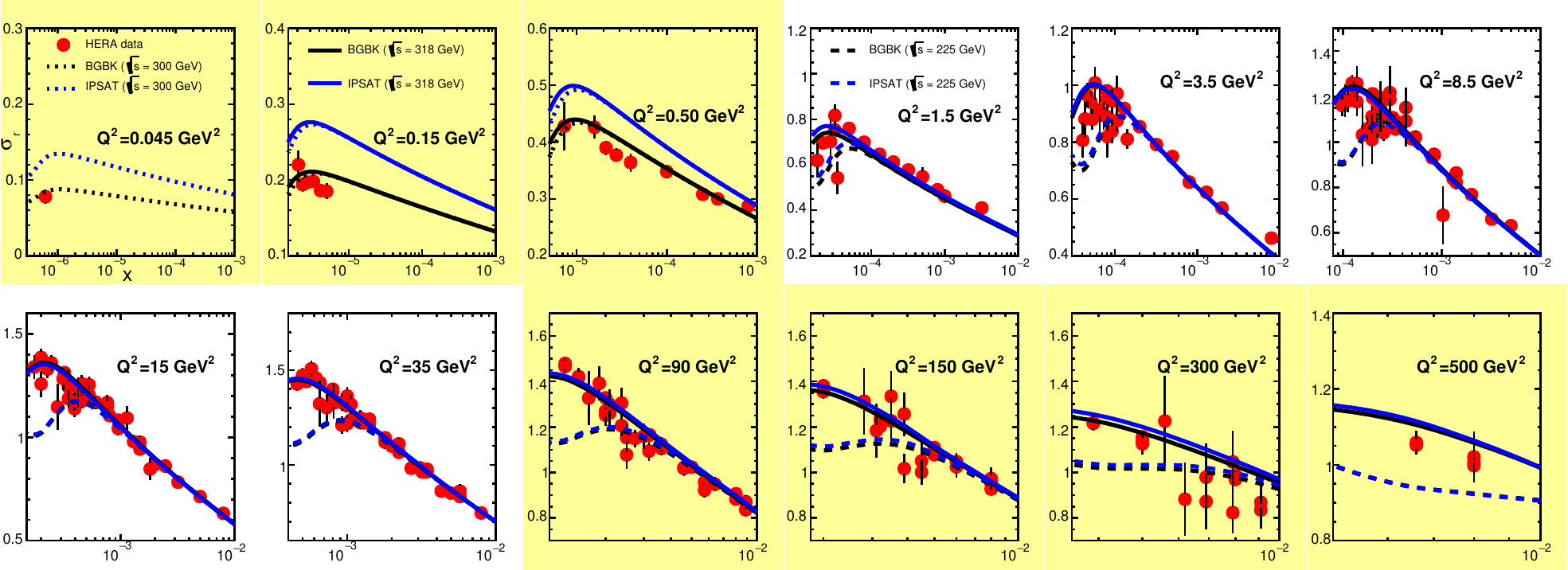}
    \caption{ \textit{White frames}: Reduced cross section, $\sigma_{r}$, fits to HERA $e^{\pm}p$ correlated data \cite{Abramowicz:2015mha} in the range $x \leqslant 0.01$ and 1.5 GeV$^{2}$ $\leqslant Q^{2} \leqslant 50$ GeV$^{2}$ with dipole models BGK and IP-sat. \textit{Yellow frames}: predictions from both models to lower and larger $Q^{2}$, relative to bin extrema. \textit{Curves} are calculated using the best-fit parameters given in Table \ref{tab:Model2_bin1} for c.m. energies comprising each $Q^{2}$ set: $\sqrt{s}=225$ GeV (dashed), 300 GeV (dotted) and 318 GeV (solid).}
    \label{fig:SigRed}
\end{figure}

In Figure \ref{fig:SigRed} one shows our results for reduced cross section from models BGK and IP-sat in the range $Q^{2}:(1.5,50)$ GeV$^2$, using the parameters of \ref{tab:Model2_bin1} corresponding to fits to dataset I,  in white frames. Predictions to lower and higher $Q^2$ values are also furnished in this plot, in the yellow frames. When appropriate, fits and predictions are given for two c.m. energies, which corresponds to two choices for the inelasticity $y$ in the reduced cross section calculations. This procedure is motivated by the fact that HERA $e^{\pm}p$ data have been taken at different energies, $\sqrt{s}$, and in this case both, fits and predictions to $\sigma_{r}$ shall accommodate different inelasticities for the pair of kinematical variables $(x,Q^{2})$.

\begin{figure}
    \centering
    \includegraphics[scale=0.85]{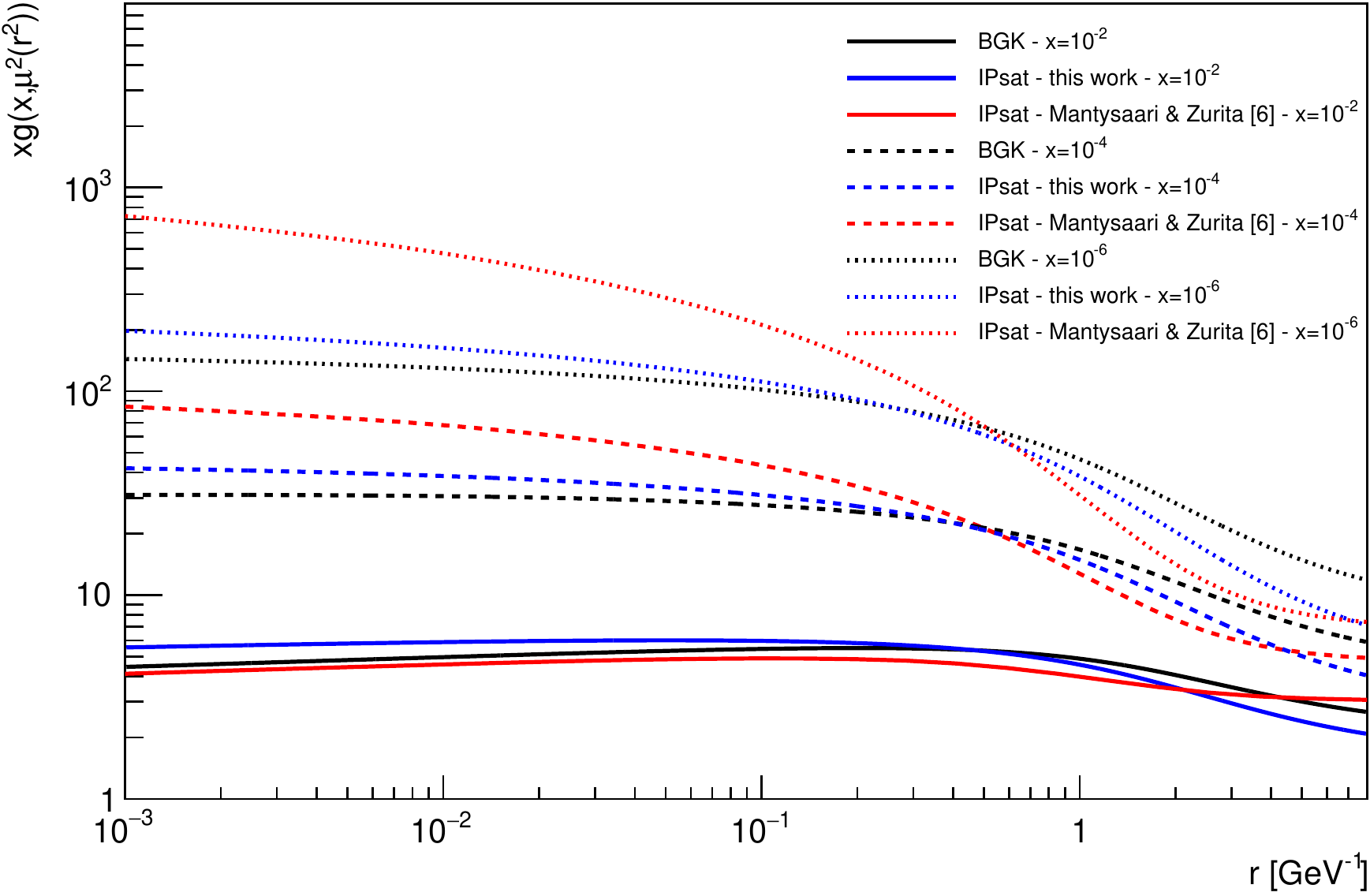}
    \caption{ Gluon density, $xg(x,\mu^{2}(r))$, as a function of the dipole transverse size, $r$, from Fit I for $x=10^{-2}, 10^{-4}$ and $10^{-6}$ for BGK and IP-sat dipole models. Our calculations with IP-sat are compared (at the same $x$ values) to the one of Ref. \cite{Mantysaari:2018nng}. }
    \label{fig:gluon_bin1_BGK}
\end{figure}

The evolution of the analytical gluon density of Eqs. (\ref{eq:gluon_initial}-\ref{eq:newSigma}) is shown in Figure \ref{fig:gluon_bin1_BGK}, with the curves derived from parameters $A_{g}$, $\lambda_{g}$ and $\mathcal{N}$ displayed in the Table \ref{tab:Model2_bin1} for Dataset I. The differences observed between models BGK and IP-sat at all dipole sizes are noticeable, though expected as the impact parameter structure is handled distinctively as previously discussed in Section \ref{sec:2}. The presence of an extra fit parameter in the BGK model also play an important role in the parameters ruling gluon evolution in our model. Such feature can be readily noticed by checking that IP-sat predictions to lower $Q^{2}$ virtualities in Figure \ref{fig:SigRed} overshoots the data. Even though, the dipole size-dependent scale $\mu^{2}(r)$ and the photon virtualy $Q^{2}$ are only linked through the wave functions in the formalism, fixing $B_{G}=4.0$ GeV$^{-2}$ in the core of IP-sat has a clear effect in the region $Q^{2}\lesssim \mu^{2}_{0}$. Additional tests varying the slope in the range $B_{G}=4.0-4.8$ GeV$^{-2}$ indicate a preference for smaller values, as one gets $\chi^{2}\sim 1.4$ for higher slopes. To settle the question of an optimal value of $B_{G}$ one should perform simultaneous fit to exclusive vector meson data, at least at same level of precision as HERA's inclusive data, or to parameterize the energy evolution of $B_{G}(W)$ of $J/\Psi$ production data \cite{Cepila:2019skb,Wang:2022jwh}. Since both alternatives go beyond the scope of the present work and we leave it for the future studies.

Moreover, despite the simplicity of the model, one finds very good agreement between our gluon density and previous published ones, such as \cite{Rezaeian:2012ji,Mantysaari:2018nng}, often obtained through numerical evolution of DGLAP equations. For instance, we observe a slow decrease of the gluon density in the large dipole domain, for $x\sim 10^{-2}$, such as revealed in Refs. \cite{Kowalski:2003hm,Mantysaari:2018nng}.

\begin{figure}[H]
    \centering
    \includegraphics[scale=0.8]{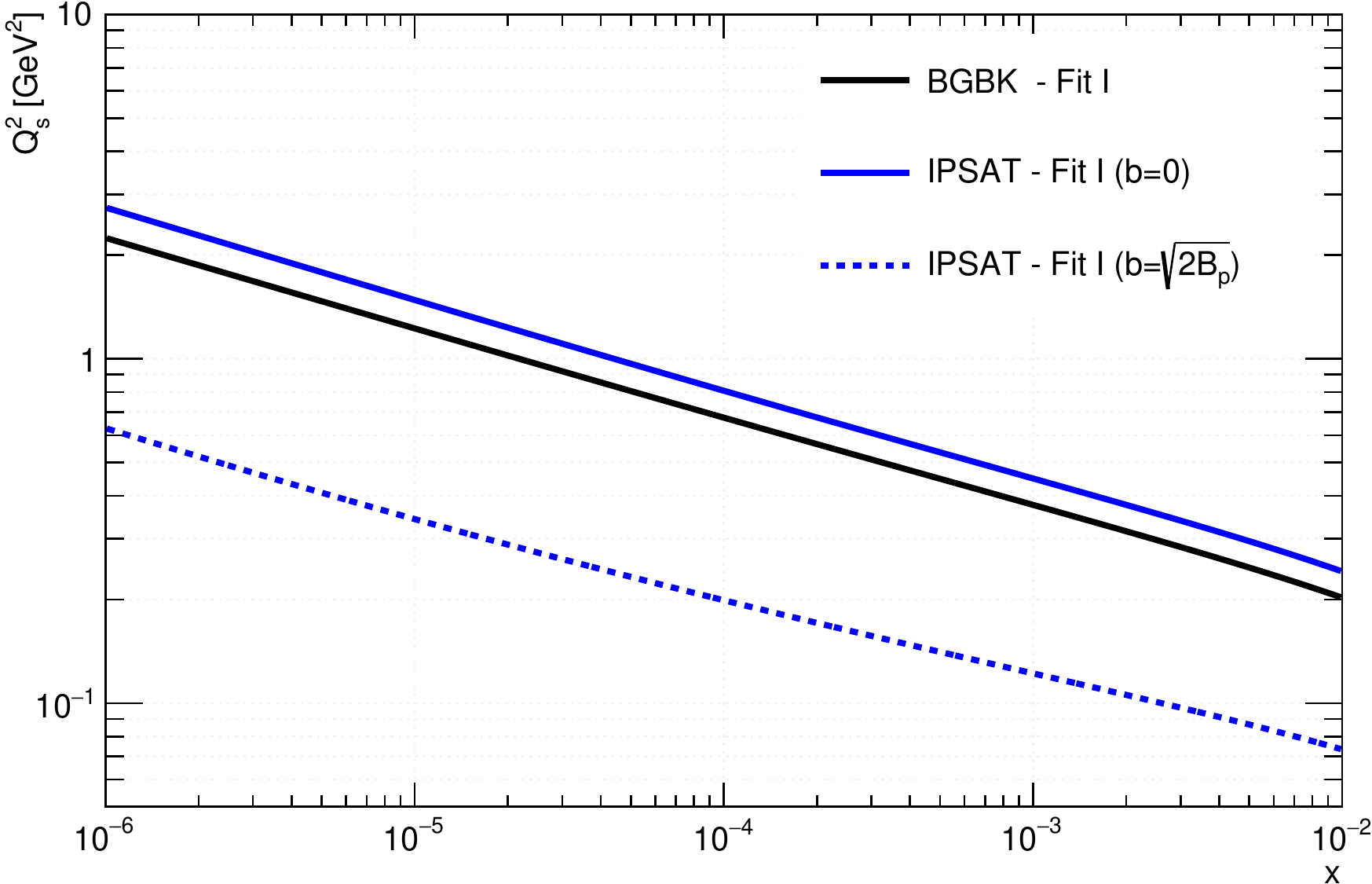}
    \caption{Saturation scale for models BGK (black solid) and IP-sat at impact parameter $b=0$ (blue solid) and the rms proton radius, $b=\sqrt{2B_{G}}\approx 2.83$ GeV$^{-1}$ (blue dashed).}
    \label{fig:Q2s}
\end{figure}

The very good fits to HERA low$-x$ data obtained in Table \ref{tab:Model2_bin1} and displayed in Figure \ref{fig:SigRed}, for $Q^{2}_{min}=1.5$ GeV$^{2}$ (i.e close to perturbative scale $\mu^{2}_{0}$), indicate that saturation effects may indeed become important at $Q^{2}\lesssim \mu^{2}_{0}$. That shall translate into a saturation scale, $Q_{s}^{2}\sim \mu^{2}_{0}$, for $x\lesssim 10^{-4}$. Indeed, with $Q_{s}^{2}(x)$ defined here as the scale for which
\begin{equation}
   \hat{\sigma}_{dip}(r^{2}=2/Q^{2}_{s}(x),x,b)=1-e^{-1/2},
\end{equation}
one obtains the results displayed in the Figure \ref{fig:Q2s}. This plot shows the variation $Q_{s}^{2}(x)$ for models BGK and for two impact parameters values in the case of IP-sat: (i) $b=0$ and (ii) at the r.m.s proton radius, $R_{p}=\sqrt{\langle b^{2} \rangle}=\sqrt{2B_{G}}=2.82$ GeV$^{-2}$. As expected one finds, following the above definition, that $Q^{2}_{s} \sim 1.0-3.0$ GeV$^{2}$ in the low$-x$  region $ 10^{-6} \lesssim  x \lesssim 10^{-5}$. Moreover, we have tested the sensitivity of the saturation scale with minimum photon virtuality in bin fits, finding very similar results even for $Q^{2}_{min}=$ 0.045 GeV$^{2}$. Once again, these findings confirm previous results using numerical DGLAP evolution within the color dipole framework (see for instance Refs.\cite{Kowalski:2003hm,Rezaeian:2012ji,Mantysaari:2018nng}).

\begin{figure}[H]
    \centering
    \includegraphics[scale=0.8]{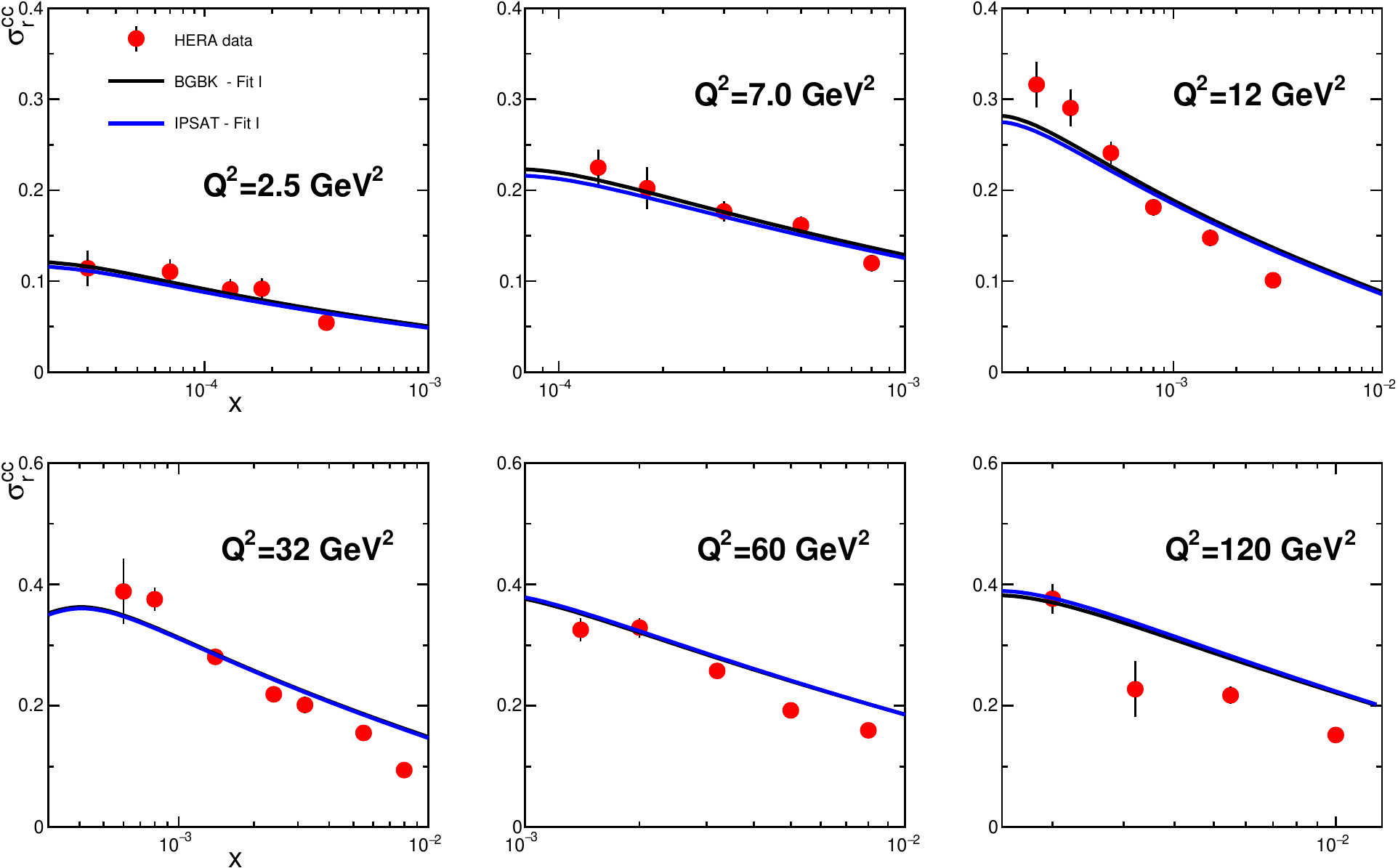}
    \caption{Charm reduced cross section data, $\sigma_{r}^{c\bar{c}}(x,Q^{2})$, from HERA in the range 2.5 GeV$^{2}$ $\leqslant Q^{2} \leqslant 120$ GeV$^{2}$ \cite{H1:2018flt}. \textit{Curves} correspond to predictions of Fit I for models BGK and IP-sat.} 
    \label{fig:SigRedcc}
\end{figure}

\begin{figure}[H]
    \centering
    \includegraphics[scale=0.8]{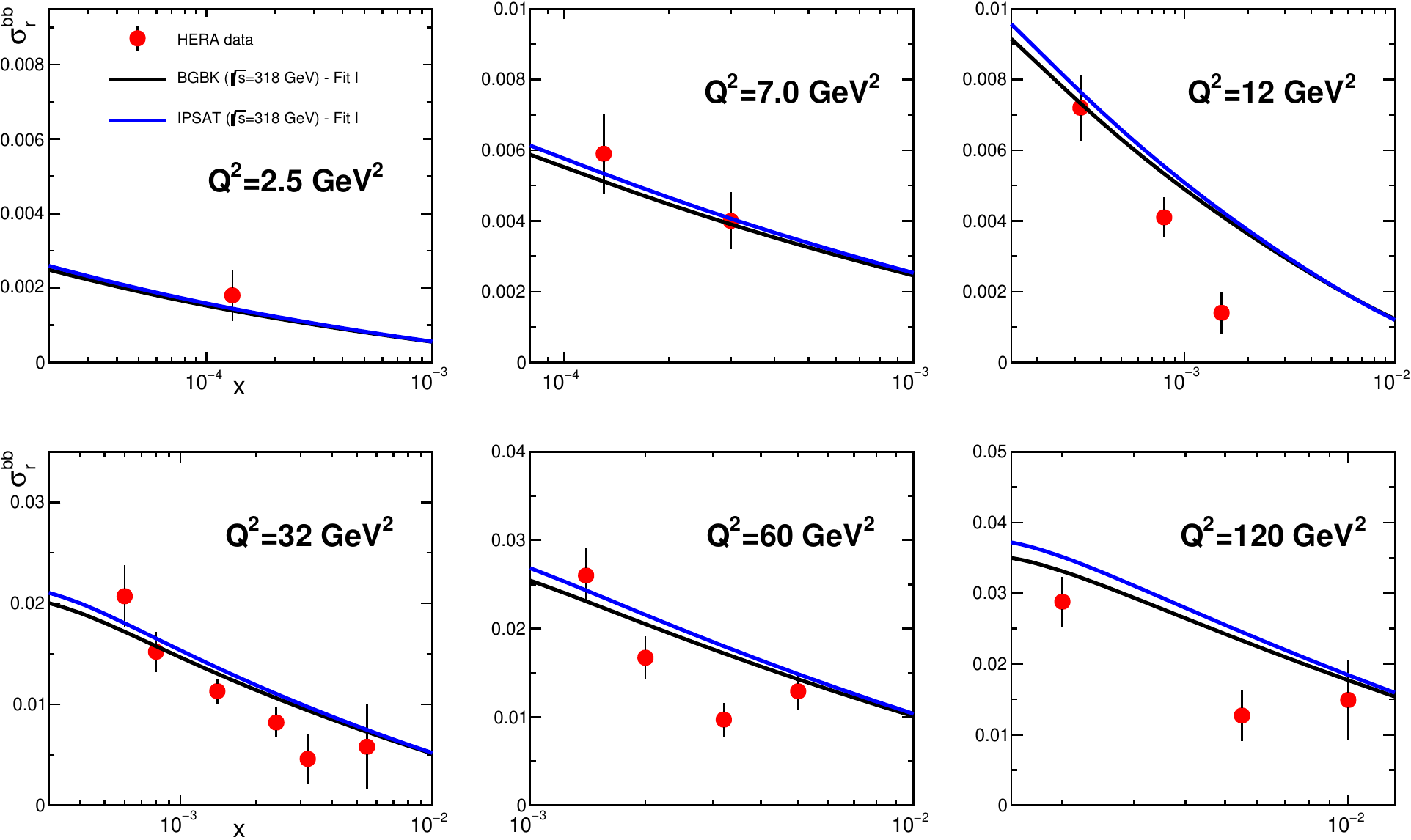}
    \caption{Bottom reduced cross section data, $\sigma_{r}^{b\bar{b}}(x,Q^{2})$, from HERA in the range 2.5 GeV$^{2}$ $\leqslant Q^{2} \leqslant 120$ GeV$^{2}$ \cite{H1:2018flt}. \textit{Curves} correspond to predictions of Fit I for models BGK and IP-sat.} 
    \label{fig:SigRedbb}
\end{figure}

In addition, heavy quark structure functions from HERA have also been analyzed in a wide $Q^2$ range, namely  $2.5 \,\,\mathrm{GeV}^2\leqslant Q^2 \leqslant 120$ GeV$^{2}$. In Fig. \ref{fig:SigRedcc} we display charm reduced cross section data, $\sigma_{r}^{c\bar{c}}$, together with BGK and IP-sat predictions from Fit I, for which no charm data have been included in the fits. Both models describe the global features of the data, specially at low$-x$ where the color dipole model is expected to hold. Moreover, we notice that adding charm data to these fits has no significant impact in their description. Such effect can be explained by two reasons: (i) a sparse charm cross section ensemble, whose fit weight is barely significant in comparison with the one of $\sigma_{r}$; (ii) fixing the charm mass at 1.3 GeV renders little flexibility to the charm wave function. The former is obviously a leading factor. The same features hold for bottom reduced cross section data, whose data and predictions are displayed in Fig. \ref{fig:SigRedbb}.

\begin{figure}[H]
    \centering
    \includegraphics[scale=0.8]{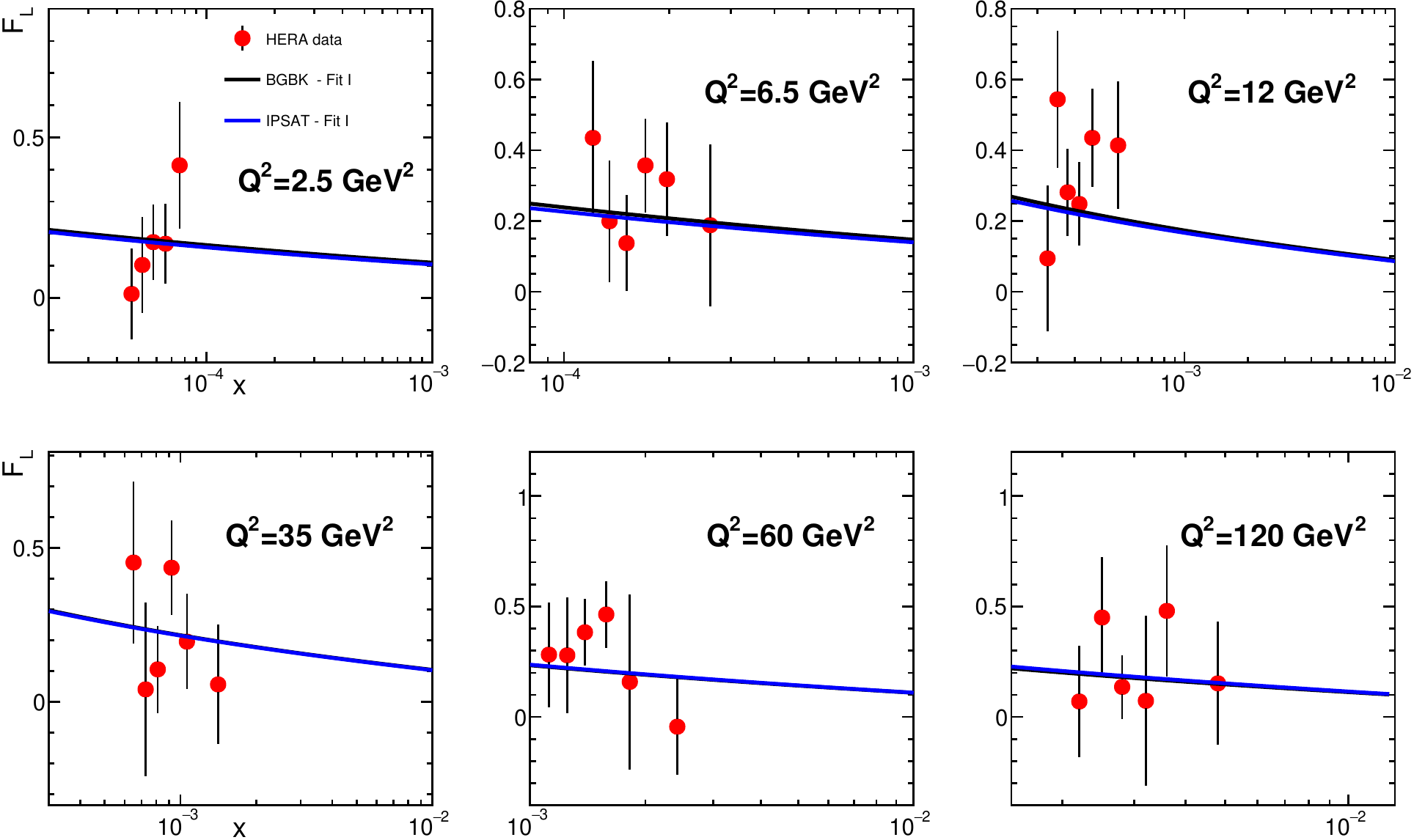}
    \caption{Longitudinal structure function, $F_{L}(x,Q^{2})$, from HERA in the range 1.5 GeV$^{2}$ $\leqslant Q^{2} \leqslant 120$ GeV$^{2}$ \cite{Andreev:2013vha} and prediction of BGK and IP-sat models for parameters obtained in Fit I.}
    \label{fig:FL}
\end{figure}

A final consistence test of our analytical gluon model is shown in the plot of Fig. \ref{fig:FL}, where one shows HERA data for the longitudinal structure function, $F_{L}$, alongside our predictions. These results show good agreement between model and data, specially considering the large uncertainties in this observable. Nevertheless, the general trend of the data is nicely reproduced by BGK and IP-sat, specially for $x\lesssim 10^{-3}$ and  30 GeV$^{2} \lesssim Q^{2} \lesssim$ 120 GeV$^{2}$. In this aspect, future measurements made at EIC and at LHeC shall be made at much lower uncertainty level and reveal a much more accurate view of longitudinal structure of the proton.

\begin{figure}[H]
    \centering
    \includegraphics[scale=0.8]{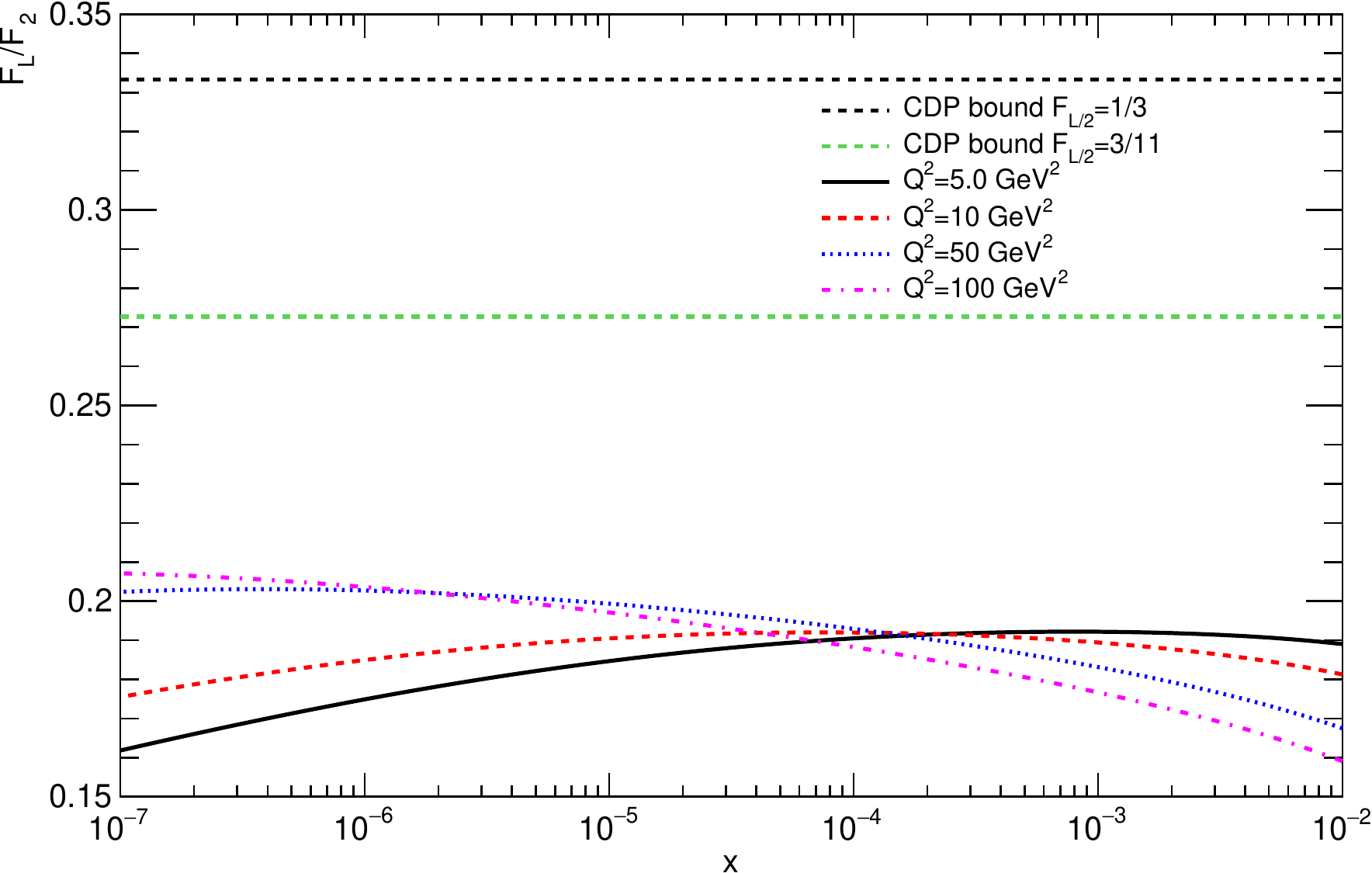}
    \caption{$Q^{2}$ evolution of the ratio $F_{L}/F_{2}$ obtained with the BGK dipole model compared to Color Dipole Picture (CDP) bounds from Ref.\cite{Boroun:2021ekf}.}
    \label{fig:FL/2 ratio}
\end{figure}

With that in mind, one shows in the Figure \ref{fig:FL/2 ratio} our predictions of the ratio $F_{L/2} \equiv F_{L}/F_{2}$ for the BGK dipole model using the gluon analytical model of Eq. (\ref{eq:model2.2}). Additionally, we furnish several predictions of both dipole models in the range $(x,Q^{2}): (10^{-6}-10^{-2}, 5.0-50\ \text{GeV}^{2})$ in Table \ref{tab:predictions}, to be probed at upcoming collider experiments such as EIC.

Moreover, the predictions given in Fig. \ref{fig:FL/2 ratio} can be directly compared to bounds of the color dipole picture, as given in Ref.\cite{Boroun:2021ekf}. This plot shows some interesting features of the ratio $F_{L/2}(x,Q^{2})$, namely: (i) while at moderate $Q^{2}$ the ratio $F_{L/2}$ is far from the achiever any of the CDP bounds, its increase with $Q^{2}$ for $x\lesssim 10^{-4}$ yields a larger fraction of $\sigma_{\gamma^{*}p}$ due to longitudinal photon polarization and (ii) the opposite behavior for $x\gtrsim 10^{-4}$, that is, transverse polarization starting to contribute more for higher $x$ and $Q^2$. These features compose an interesting tool to investigate the longitudinal structure of the proton at upcoming colliders such EIC, as they are expected to improve the precision of $F_{L}$ measurements at low$-x$ and moderate $Q^{2}$ \cite{Badelek:2022cgr}. 

{It is worth mentioning that the gluon distribution of the proton at low-$x$ has been recently determined in the context of the color dipole picture in Ref. \cite{Boroun:2022uot}. It is expressed in terms of the structure function $F_2$ and  the ratio $R=F_L/F_2$. At large $Q^2$, it is given by:
\begin{eqnarray}
xg(x,\mu^2) = \frac{9\pi}{\alpha_s(\mu^2)N_c(\sum_f e_f^2)}\frac{F_2(\xi_L x, \mu^2) }{1+R^{-1}} \approx N_g\left(\frac{Q^2}{1\,\mathrm{GeV^2}}\frac{1}{\xi_Lx}  \right)^{0.29},
\end{eqnarray}
where $e_f^2$ are the squared charges of the active flavours, $\xi_L\simeq 0.40$ is the rescaling factor and $N_g$ the overall normalization. The last result is based on a two-parameter eye-ball fit to the experimental data, $F_2(W^2) = f_2(W^2/\mathrm{Gev^2})^{C_2} $ with $f_2=0.063$ and $C_2=0.29$. It represents the asymptotic representation of the full calculation shown in Ref. \cite{Boroun:2022uot} and it provides a good approximation for $Q^2>30$ GeV$^2$. At $Q_0^2=1.9$ GeV$^2$, the results of \cite{Boroun:2022uot} can be parametrized as $xg(x,Q_0^2)\approx 0.5\,x^{-0.21}(1-x)^6$ which is not so far from the fits presented in Table \ref{tab:Model2_bin1}. }

\begin{table}[H]
    \centering
    \caption{Predictions of structure functions $F_{2},F_{L}$ and heavy quark structure functions, $F_{2}^{c\bar{c}},F_{2}^{b\bar{b}}$ of models BGK and IP-sat using the gluon pdf of eq.(\ref{eq:model2.2}).}
    \vspace{.3cm}
\resizebox{.5\textwidth}{!}{
    \begin{tabular}{c|c|c|c|c|c|c}
    \thickhline
         Model & $x$ & $Q^2$ [GeV$^{2}$] & $F_{2}$ & $F_{L}$ & $F_{2}^{c\bar{c}}$ & $F_{2}^{b\bar{b}}$ \\
         \hline
         \multirow{9}{*}{BGK}  & \multirow{3}{*}{$10^{-2}$} & 5.0 & 0.430 & 0.076 & 0.043 & 0.00019 \\
         \cline{3-7}
         &  & 10 & 0.524 & 0.087 & 0.078 & 0.00088 \\
        \cline{3-7}
         &  & 50 & 0.749 & 0.108 & 0.175 & 0.0084 \\
        \cline{2-7}
         & \multirow{3}{*}{$10^{-4}$} & 5.0 & 1.134 & 0.217 & 0.175&0.0037  \\
        \cline{3-7}
         &  & 10 & 1.497 & 0.277 &  0.300 & 0.0085 \\
         \cline{3-7}
         &  & 50 & 2.467  & 0.421 & 0.690 & 0.046 \\
         \cline{2-7}
         & \multirow{3}{*}{$10^{-6}$} & 5.0 & 2.452 & 0.475 & 0.476  & 0.013\\
         \cline{3-7}
         &  & 10 & 3.490 & 0.674 & 0.829 & 0.030 \\
         \cline{3-7}
         &  & 50 & 6.611 & 1.241 & 2.037& 0.155\\
         \hline
         \multirow{9}{*}{IP-sat}  & \multirow{3}{*}{$10^{-2}$} & 5.0 &  0.427 & 0.072 & 0.041 & 0.00018 \\
         \cline{3-7}
         &  & 10 & 0.521 & 0.083 & 0.075 & 0.00086 \\
         \cline{3-7}
         &  & 50 & 0.751 & 0.108 & 0.175 & 0.0086\\
        \cline{2-7}
         & \multirow{3}{*}{$10^{-4}$} & 5.0 & 1.125 & 0.205 & 0.169 & 0.0038\\
         \cline{3-7}
         &  & 10 & 1.480 & 0.264 & 0.292 & 0.0089 \\
         \cline{3-7}
         &  & 50 & 2.463 & 0.420 & 0.690 & 0.048\\
     \cline{2-7}
         & \multirow{3}{*}{$10^{-6}$} & 5.0 & 2.474 & 0.462 & 0.456 & 0.014 \\
        \cline{3-7}
         &  & 10 & 3.466 & 0.647 & 0.798 & 0.031 \\
         \cline{3-7}
         &  & 50 & 6.565 & 1.216 & 2.001 & 0.160 \\
    \thickhline
    \end{tabular}
    }
    \label{tab:predictions}
\end{table}

\section{Exclusive diffractive photoproduction }
\label{sec:4}
{The differential cross section of exclusive vector meson photoproduction $\gamma p \rightarrow Vp$ is given by} \cite{Kowalski:2006hc}:
\begin{eqnarray}
  \frac{d\sigma^{\gamma p\rightarrow Vp}}{d t} = \frac{1}{16\pi}\left|\mathcal{A}^{\gamma p\rightarrow Vp}\right|^2\;(1+\beta^2)\,R_g^2,
  \label{exclusive-meson}
\end{eqnarray}
{where the real-to-imaginary ratio of the scattering amplitude, $\beta$, is computed by using}
\begin{eqnarray}
  \beta = \tan\left(\frac{\pi\lambda_{\mathrm{eff}}}{2}\right), \quad\text{with}\quad \lambda_{\mathrm{eff}} \equiv \frac{\partial\ln\left(\mathcal{A}_{T}^{\gamma p\rightarrow Vp}\right)}{\partial\ln(1/x)},
\end{eqnarray}
{and the skewedness factor $R_g^2$ is calculated using}
\begin{eqnarray} 
  R_g(\lambda_{\mathrm{eff}}) = \frac{2^{2\lambda_{\mathrm{eff}}+3}}{\sqrt{\pi}}\frac{\Gamma(\lambda_{\mathrm{eff}}+5/2)}{\Gamma(\lambda_{\mathrm{eff}}+4)}.
\end{eqnarray}

{Within color dipole framework, the elastic scattering amplitude for the process $\gamma+p\rightarrow V+p$ is a function of $x$ and of the momentum transfer $\vec{\Delta}$ (with $|t|=\vec{\Delta}^2$), being written as a Fourier transform of the photon and vector meson wavefunctions convoluted with the color dipole scattering amplitude \cite{Kowalski:2006hc}:}
\begin{eqnarray} 
  \mathcal{A}^{\gamma p\rightarrow Vp} & = & \mathrm{i}\,\int\!d^2\vec{r}\int_0^1\!\frac{d{z}}{4\pi}\int\!d^2\vec{b}\;(\Psi_{V}^{*}\Psi_{\gamma})_{T}\;\mathrm{e}^{-\mathrm{i}[\vec{b}-(1-z)\vec{r}]\cdot\vec{\Delta}}\;\frac{d\sigma_{q\bar q}}{d^2\vec{b}}, \\
  & = &  \mathrm{i}\,\pi\int_0^\infty\!r\,d{r}\,\int_0^1d{z}\int_0^\infty\!b\,d{b}\,(\Psi_V^*\Psi_{\gamma})_{T}\;J_0(b\Delta)\;J_0\left([1-z]r\Delta\right)\;\frac{d\sigma_{q\bar{q}}}{d^2\vec{b}}, 
\end{eqnarray}
{where $J_{0}(x)$ is the zeroth order Bessel function of the first kind.}

{The overlap between the transverse\footnote{As we are  interested here in the photoproduction limit of vector mesons, $J/\psi$ and  $\rho$, only the transverse part of the product ($\Psi^{*}_{V}\Psi_{\gamma}$) is relevant.} photon and the vector meson wave functions is given by:}
\begin{eqnarray}
  (\Psi_V^*\Psi_{\gamma})_{T} = \hat{e}_f \sqrt{4\pi\alpha_{em}}\, \frac{N_c}{\pi z(1-z)} \,
  \left\{m_f^2 K_0(\epsilon r)\phi_T(r,z) - \left[z^2+(1-z)^2\right]\epsilon K_1(\epsilon r) \partial_r \phi_T(r,z)\right\},
\end{eqnarray}
{where the effective charge $\hat{e}_f=2/3$ and $1/\sqrt{2}$, for $J/\psi$ and  $\rho$ mesons, respectively.} 
{In this study the \textit{boosted Gaussian} wave function \cite{Kowalski:2006hc} is considered, for which the scalar part of meson wave function $\phi_{T,L}$ is expressed as,}
\begin{eqnarray}
  \phi_{T,L}(r,z) = \mathcal{N}_{T,L} z(1-z)
  \exp\left(-\frac{m_f^2 \mathcal{R}^2}{8z(1-z)} - \frac{2z(1-z)r^2}{\mathcal{R}^2} + \frac{m_f^2\mathcal{R}^2}{2}\right),
\end{eqnarray}
{where the corresponding parameters $\mathcal{N}_{T,L}$ and $\mathcal{R}$ are properly obtained from both wave function normalization and constraint from the electronic decay width, $\Gamma_{V\to e^+e^-}$. The obtained values using the quark masses of $m_f=0.03$ GeV for $\rho$
 and  $m_f=1.3$ GeV for $J/\psi$ are presented in Table \ref{tab:BG_params}. The predicted and measured decay width are also presented.}

\begin{table}[H]
    \centering
    \caption{Parameters of boosted Gaussian wave function of vector mesons $J/\psi$ and $\rho$ obtained for quark masses tunned by fits to $\sigma_{r}$ in Table \ref{tab:Model2_bin1} for dipole models BGK and IP-sat.}
    \begin{tabular}{c|c|c|c|c|c|c|c}
    \thickhline
         \text{Meson} & $M_{V}$ [GeV]& $m_{f}$[GeV] & $\mathcal{N}_{T}$ & $\mathcal{N}_{L}$ & $\mathcal{R}$ [GeV$^{-1}$] & $\Gamma_{V\rightarrow e^{+}e^{-}}^{\text{exp}}$ [keV] & $\Gamma_{V\rightarrow e^{+}e^{-}}^{\text{calc}}$ [keV] \\
         \hline
         $J/\psi $ & 3.097 & 1.3 & 0.5974 & 0.5940 & 1.5181 &  $5.53 \pm 0.11 $ & 5.53 \\
         \hline
         $\rho $ &0.7753 & 0.030 & 0.9942  & 0.8928 & 3.6388 & $ 7.04\pm 0.06$ & 7.04 \\
      \thickhline
    \end{tabular}

    \label{tab:BG_params}
\end{table}
\noindent

\begin{figure}[H]
    \centering
    \includegraphics[scale=0.8]{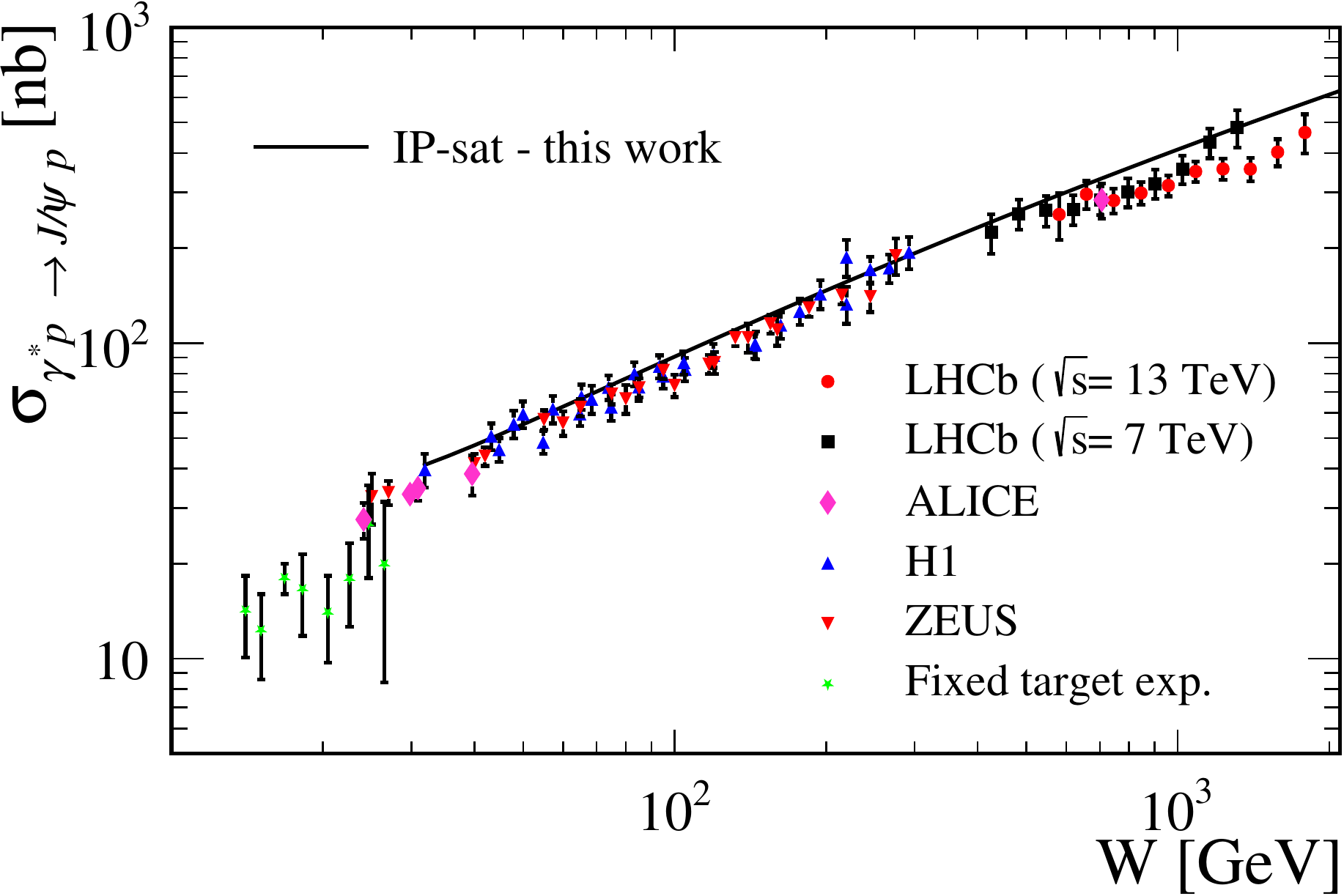}
    \caption{$J/\psi$ photoproduction cross section data (compiled in Ref. \cite{LHCb:2018rcm}) and prediction of the IP-sat dipole model with asymptotic gluon density of eq. (\ref{eq:model2.2}) and parameters from fit I of Table \ref{tab:Model2_bin1}.}
    \label{fig:JPsiXsec}
\end{figure}

{The result for $J/\psi$ photoproduction is presented in Fig. \ref{fig:JPsiXsec} for the IP-sat model and the fitted gluon distribution presented in Eq. (\ref{eq:model2.2}). For comparison, we display with our predictions alongside DESY-HERA data, fixed target experiments and extracted cross sections from proton-proton ultraperipheral collisions as well (LHCb and ALICE data) (with the data gathered Ref. \cite{LHCb:2018rcm}.}). As it can be seen from this plot, the overall normalization and energy dependence are correctly described. Moreover, we recall that this is a parameter-free prediction as the dipole cross section is determined from parameters of Fit I (see Table \ref{tab:Model2_bin1}). 

\begin{figure}[H]
    \centering
    \includegraphics[scale=0.8]{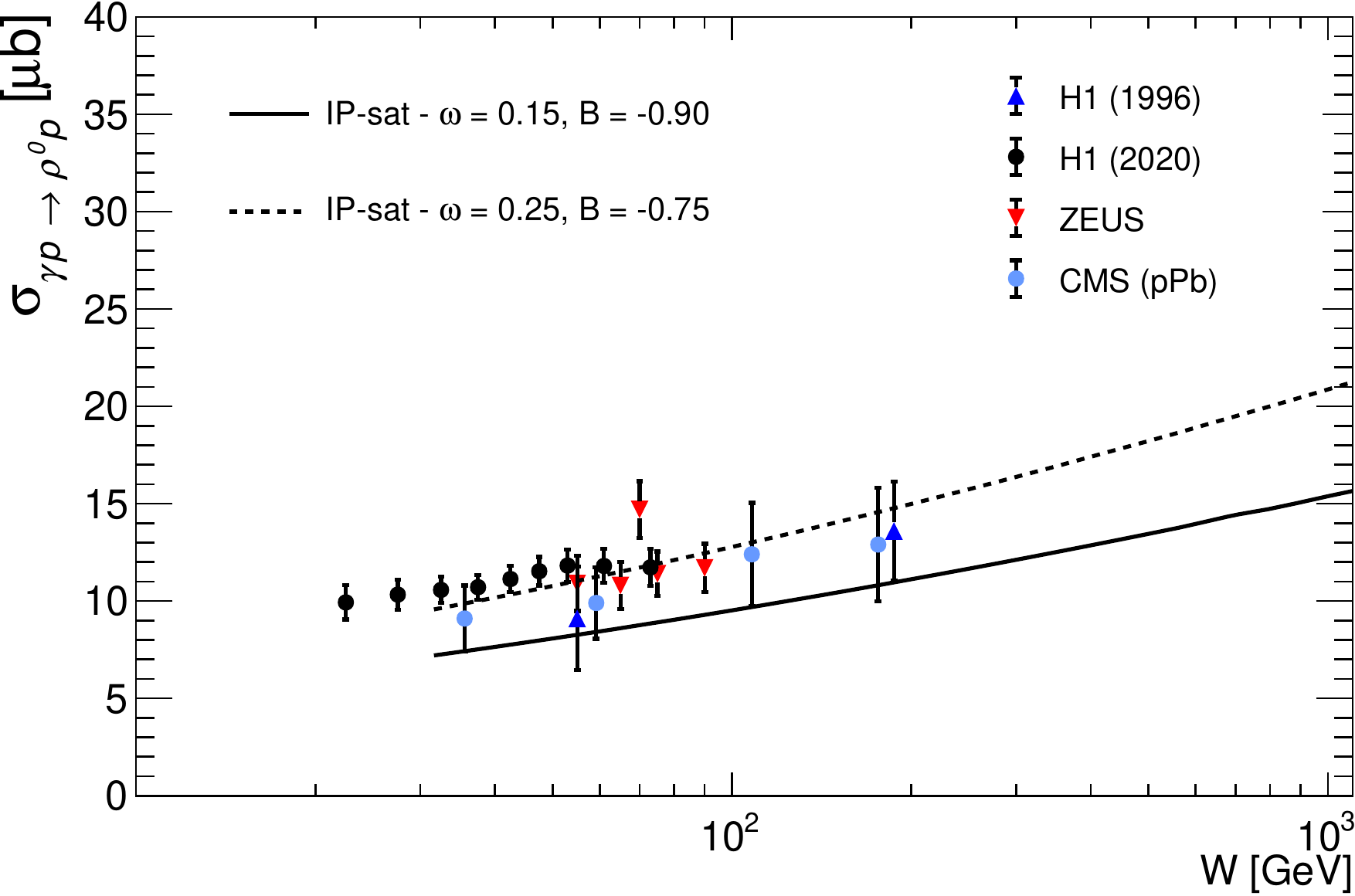}
    \caption{$\rho(770)$ photoproduction cross section data and prediction of the IP-sat dipole model with asymptotic gluon density of eq. (\ref{eq:model2.2}) and parameters from fit I of Table \ref{tab:Model2_bin1}. The nonperturbative correction of the meson wave function, $f_{s}$, in eq. (\ref{factor-fs}) is evaluated at $B=-0.90$, $\omega=0.15$ (as suggested in Ref. \cite{Goncalves:2020cir}). and the additional set, $B=-0.75$, $\omega=0.25$, provides a better fit to the data. }
    \label{fig:RhoXsec}
\end{figure}

{The Fig. \ref{fig:RhoXsec} comprises our results for $\rho$ photoproduction, also compared to DESY-HERA data and the  CMS Collaboration (extracted from ultraperipheral pPb collisions). Here, some discussion is in order. In the $\rho$ production at $Q^2=0$ no hard scale is present and the photon wave function is dominated by large transverse  dipole size contributions. Therefore, confinement effects should be significant and a pragmatic approach modifies the photon wave function in order to take into account such a non-perturbative (soft) contribution. Following Ref. \cite{Frankfurt:1997zk,Forshaw:1999uf}, the photon wave function is replaced by}
\begin{eqnarray}
  \psi_{T,L}^f(Q,r,z)\rightarrow \sqrt{f_s(r)}\,\psi_{T,L}^f(Q,r,z), \quad f_{s}(r) = \left[
 \frac{1 + B \exp\left( -\omega^{2} (r -  R)^{2} \right)}
 {1 + B \exp\left( -\omega^{2} R^{2} \right)}
 \right],
 \label{factor-fs}
\end{eqnarray}
{where the parameters $B$, $\omega$ and $R$ are determined by fitting the total  photoproduction cross section, $\sigma (\gamma p\rightarrow X)$. The shifted
Gaussian function, $f_s(r)$, above controls the width and height of the soft contribution enhancement for the photon wavefunction. Moreover, $f_s\rightarrow 1$ for small dipoles and the hard contribution is unchanged. The original values, determined in Ref. \cite{Forshaw:1999uf} are $B= 6.8\pm 0.1$, $\omega = 0.342 \pm 0.008
$ and $R = 5.67 \pm 0.03$ (fit including charm contribution, GBW dipole cross section). More recently, these parameters have been qualitatively extracted  in the analysis on Ref. \cite{Goncalves:2020cir} obtaining  $B = -0.90$ and $\omega = 0.15$ GeV with fixed $R = 6.8$ GeV$^{-1}$. The prediction using these values is represented by the solid line in Fig. \ref{fig:RhoXsec}. With this set of parameters the data are underestimated by a factor $\simeq 0.8$. We have tested another values and the optimal set of parameters for the IP-SAT model with fit I is  $B = -0.75$, $\omega = 0.25$ GeV and $R = 6.8$ GeV$^{-1}$ (fixed). The result is represented by the dashed curve which describes the  normalization and shape of the $\rho$ photoproduction cross section. It should be noticed that the values of the parameters in $f_s$ function  are model dependent and correlated to the limit of the dipole cross section for a given model at large $r$.}



\section{Conclusions}
\label{sec:conc}
In this work we analyzed the applicability of a new analytical gluon density, inspired by the DAS solution obtained long ago by Ball and Forte \cite{Ball:1994du}, within the color dipole approach to the highest precision HERA data. Specifically, we have tested two dipole models, IP-sat and BGK, with and without impact parameter dependence, obtaining very good fits to HERA data in the range 1.5 GeV$^{2}\leqslant Q^{2}\leqslant 50$ GeV$^{2}$, for both models, visually and in statistical terms. While the influence of the impact parameter structure cannot be perceived from the fits in this kinematic window, the extrapolation to lower and higher $Q^{2}$ allows a good discrimination between the models IP-sat and BGK, as long as only the inclusive reduced cross section, $\sigma_{r}$, is regarded. In particular, one shows a excellent agreement  with data for BGK model in the wide range  0.045 GeV$^{2}\leqslant Q^{2}\leqslant 500$ GeV$^{2}$. On the other hand, IP-sat results hint at a better suit for higher photon virtualities. Nonetheless, we furnished also the predictions BGK and IP-sat following from our fits to other observables measured at HERA, such as $\sigma_{r}^{c\bar{c}}$, $\sigma_{r}^{b\bar{b}}$ and $F_{L}$, finding for all good agreement with datasets in the range  2.5 GeV$^{2}\leqslant Q^{2}\leqslant 120$ GeV$^{2}$.

The evolution of our gluon density with the dipole size, $r$, is shown to be pretty similar to previous studies reported in the literature (see e.g. Refs. \cite{Rezaeian:2013tka,Mantysaari:2018nng}) using numerical DGLAP evolution of the gluon within the dipole amplitude. In the scope of dipole models, our study contributes to an easier modelling of the gluon and of the dipole cross section, as the simple analytical formula one gives requires very little computational power to calculate inclusive and exclusive DIS observables. Moreover, the analysis of saturation effects, specially on the saturation scale, $Q_{s}^{2}$, also show a clear agreement with previous investigations \cite{Kowalski:2003hm,Rezaeian:2012ji,Mantysaari:2018nng} and yields $Q_{s}^2\simeq 2-3$ GeV$^{2}$ at $x=10^{-6}$. {In the context of vector meson production, where dipole models have been extensively applied, we present our calculations to photoproduction cross sections of mesons $J/\psi$ and $\rho$, which correctly describe the HERA data and the recent measurements at LHCb}. Altogether, these results stress the potential applicability of the analytical gluon of eq.(\ref{eq:model2.2}) in other studies of diffractive DIS (DDIS) and exclusive particle production, such as the deeply virtual Compton scattering (DVCS), which we leave for a future work, as the focus of this paper was to provide an analytical gluon density suitable for the high-precision HERA data and to study high-energy phenomena at future colliders such as EIC and the LHeC.

\begin{acknowledgments}
We thank H. Mantys\"{a}ari for sharing with us his IP-sat code and for instructions on how to handle it.This work was supported by the Brazilian funding agencies CAPES and CNPq. DAF acknowledges the support of the project INCT-FNA (464898/2014-5).
\end{acknowledgments}

\appendix
\section{DAS gluon distribution including heavy quark contribution}
{In this section we discuss the results of fits with the DAS model including heavy quarks for the BGK model. In the analysis the following masses were used: $m_{lq}=0.03$ GeV, $m_c=1.3$ GeV and $m_b=4.2$ GeV, respectively. The results of the fits with the small-$x$ data are summarized in Table \ref{tab:Model1_bin1_hqs}. In the first two rows we show the values of $\chi^2/\mathrm{dof}$ and $p$-value, obtained with the fixed parameters $x_0$ and $\mu_0^2$  following the values obtained in the light quark fit in Table \ref{tab:Model1_bin1}.  Rather large
values of $\chi/\mathrm{dof}$ are obtained for $x_0=1$ whereas for $x_0=0.1$ the quality of fit is improved. However, a still large value of $\sigma_0$ is found suggesting the need of more flexible fit (including new free parameters). In the next two rows, results are presented by allowing the parameter $x_0$ to be free for fixed $\mu_0^2=1.1$ GeV$^2$ or considering both $x_0$ and  $\mu_0^2$ as free parameters. A good fit quality is achieved, with a number of
fitted parameters similar to the ones in Refs. \cite{Luszczak:2016bxd,Golec-Biernat:2017lfv}. The new parameters present smaller values of $x_0\sim 10^{-2}$  and $\mu_0^2$ increased by around 17$\%$. The important point is that the $\sigma_0$ has diminished but still high compared to $\sigma_0\simeq 30$ mb. On the other hand, values $\sigma_0\simeq 100$ mb are also obtained in fits using BGK model  with DGLAP evolution at NLO \cite{Luszczak:2016bxd}. We recall that in the later case, at the initial scale $\mu^{2}_{0}$ the gluon acquires another x-dependence, namely $xg(x,\mu^{2}\approx \mu^{2}_{0}) \propto \left[ \ln\left( 1/x\right)\right]^{-1/4}$ which justifies the increase of $A_{g}$ in these fits. Therefore, our choice for the modified gluon of eqs.(\ref{eq:model2.2}-\ref{eq:newSigma}) lies in the choice one makes for a \textit{soft} ansatz (following good results of Refs. \cite{Rezaeian:2012ji,Luszczak:2016bxd,Mantysaari:2018nng} in which DGLAP equations are evolved numerically) and the similar statistical results in fits row of Table \ref{tab:Model2_bin1} (compared to the fifth row of Table \ref{tab:Model1_bin1_hqs}).}

\begin{table}[H]
\centering
\caption{Fit parameters of DAS model of Eq.(\ref{eq:model1})  for $Q^{2}:(1.5,50)$ GeV$^2$ and $x\leqslant 0.001$, including charm and bottom quarks. All fit parameters are given within $70\%$ of confidence level, with $m_{lq}=0.03$ GeV, $m_{c}=1.3$ GeV and $m_{b}=4.2$ GeV, and $C=4.0$ fixed throughout.}
\vspace{.3cm}
\resizebox{.9\textwidth}{!}{
\begin{tabular}{c|c|c|c|c|c|c}
\thickhline
Model & $\sigma_0$ [mb]  & $A_{g}$ & $x_{0}$ & $\mu_{0}^{2}$ [GeV$^{2}$] & $\chi^2/$dof & $p-$value\\
\thickhline
\multirow{4}{*}{BGK} &  $ (172.4\pm 2.9)\times 10^{5} $  & $  1.624\pm 0.0041 $  & 1.0 (fixed) & 1.1 (fixed) &  $1485.02/288=5.16  $ &  0  \\
\cline{2-7}
 &  $ 295 \pm 29$  & $ 2.834 \pm 0.017 $  & 0.1 (fixed) & 1.1 (fixed) &  $306.848/288= 1.07 $ &  0.213  \\
 \cline{2-7}
& $194 \pm 17$  & $3.047 \pm 0.074$  & $0.0805\pm 0.0078 $  &  1.1 (fixed) &  $303.262/287=1.06$ &  0.244  \\
\cline{2-7}
& $114 \pm 11$  & $3.80 \pm 0.32$  & $0.0496 \pm 0.0095$  &  $1.29 \pm 0.13 $ &  $290.576/286=1.02$ & 0.414\\   
\thickhline
\end{tabular}
}
\label{tab:Model1_bin1_hqs}
\end{table}

\bibliographystyle{h-physrev}
\bibliography{analytic-gluon-fit}

\end{document}